\documentclass[secnumarabic,amssymb,amsmath,aps,prd,nofootinbib]{revtex4}

\newcommand{\s}{k}
\newcommand{\q}{n}
\newcommand{\z}{\,{\ln r}}
\newcommand{\ce}{\chi_{{}_{E}}}
\newcommand{\rrho}{R}
\newcommand{\nt}{T}
\newcommand{\ttau}{\tau}
\newcommand{\eff}{{\rm eff}}
\newcommand{\mmu}{\nu}
\newcommand{\mbr}{\mu}

\newcommand{\uu}{{\cal U}}
\newcommand{\la}{\langle}
\newcommand{\ra}{\rangle}
\newcommand{\tmn}{\la T_{\mu\nu}\ra}
\newcommand{\phisq}{\la \phi^2 \ra}
\newcommand{\kk}{\textsc{kk}}
\newcommand{\bs}{{bs}}

\usepackage{graphicx}

\begin{document}

\title{Massless scalar fields and infrared divergences
    \\in the inflationary brane world}

\author{ Oriol  Pujol{\`a}s$^1$\footnote{pujolas@yukawa.kyoto-u.ac.jp} and Takahiro
  Tanaka$^{2}$\footnote{tama@scphys.kyoto-u.ac.jp}}

\address{$^1$ Yukawa Institute for Theoretical Physics, Kyoto University, Kyoto
  606-8502, Japan}
\address{$^2$ Department of Physics, Kyoto University, Kyoto 606-8502, Japan}

\begin{abstract}

We study the quantum effects induced by bulk scalar fields in a
model with a de Sitter (dS) brane in a flat bulk (the
Vilenkin-Ipser-Sikivie model) in more than four dimensions. In
ordinary dS space, it is well known that the stress tensor in the
dS invariant vacuum for an effectively massless scalar
($m_\eff^2=m^2+\xi {\cal R}=0$ with ${\cal R}$ the Ricci scalar)
is infrared divergent except for the minimally coupled case. The
usual procedure to tame this divergence is to replace the
dS invariant vacuum by the Allen Follaci (AF) vacuum. The resulting
stress tensor breaks dS symmetry but is regular. Similarly, in the
brane world context, we find that the dS invariant vacuum
generates $\tmn$ divergent everywhere when the lowest lying mode
becomes massless except for massless minimal coupling case. A
simple extension of the AF vacuum to the present case avoids this
global divergence, but $\tmn$ remains to be divergent along a
timelike axis in the bulk. In this case, singularities also appear
along the light cone emanating from the origin in the bulk,
although they are so mild that $\tmn$ stays finite except for
non-minimal coupling cases in four or six dimensions. We discuss
implications of these results for bulk inflaton models. We also
study the evolution of the field perturbations in dS brane world.
We find that perturbations grow linearly with time on the brane,
as in the case of ordinary dS space. In the bulk, they are
asymptotically bounded.

\hfill \vbox{ \hbox{YITP-04-38, KUNS-1927} }
\end{abstract}

\maketitle

\section{introduction}

The brane world (BW) scenario \cite{rsI,rsII} has been intensively
studied in the recent years. Little is known yet concerning the
quantum effects from bulk fields in cosmological models
\cite{wade,nojiri,kks,hs}. Quite generically, one expects that
local quantities like $\tmn$ or $\phisq$ can be large close to the
branes, due to the well known divergences appearing in Casimir
energy density computations. This has been confirmed for example
in \cite{knapman,romeosaharian} for flat branes. These
divergences are of ultraviolet (UV) nature and
do not contribute to the force.
Hence, they are ignored in Casimir force computations.
However, they are relevant to the BW scenario since
they may induce large backreaction, and are worth of investigation.

In this article, we shall shed light on another aspect of
objects like $\tmn$ in BW. We shall point out that they
can suffer from infrared (IR) divergences as well.
These divergences arise when there is a zero mode in the
spectrum of bulk fields in brane models of RSII type
with dS brane\cite{rsII,gasa}.
The situation is analogous to the case in dS space
without brane.
It is well known that light scalars in dS develop an IR divergence
in the dS invariant vacuum.
The main purpose of this article is to explore the effects of
scalar fields with light modes in a BW cosmological setup of the
RSII type \cite{rsII}.
To consider massless limit of scalar field in inflating BW
is especially well motivated in the context of
`bulk inflaton' models \cite{kks,hs,shs,hts,koyama}, in which
the dynamics of a bulk scalar drives inflation on the brane. In
the simplest realizations, the brane geometry is close to dS and
the bulk scalar is nearly massless.

Let us recall what happens in the usual dS case \cite{bida}. For
light scalars $m_\eff \ll H$ (with $H$ the Hubble constant) in dS,
$\la\phi^2\ra$ and $\tmn$ in the dS invariant vacuum develop a
global IR divergence $\sim1/m_\eff^2$. To be precise, this depends
on whether the field is minimally coupled or not. What we have in
mind is a generic situation in which the effective mass
$m_\eff^2=m^2+\xi {\cal R}$ is small, and $\xi\neq0$. In these cases
$\tmn$ diverges as mentioned. The point is that in the generic
massless limit, another vacuum must be chosen to avoid the global IR
divergence. This process breaks dS invariance \cite{af}, but this
shall not really bother us. The simplest choice is the Allen Follaci
(AF) vacuum, in which the stress tensor is globally finite and
everywhere regular. The massless minimally coupled case is special
\cite{gaki}, and it accepts a different treatment which gives finite
$\tmn$ without violating dS invariance.
%, basically because
%$\la\phi^2\ra$ grows linearly with time $\chi$.

In the BW scenario \cite{rsII}, the bulk scalar is decomposed into a
continuum of KK modes and bound states. Here we consider the case
that there is a unique bound state with mass $m_d$. If $m_d$ is
light, $\la\phi^2\ra$ and $\tmn$ for the dS invariant vacuum will
also diverge like $1/m_d^2$. In this case, again, one will be forced
to take another vacuum state like the AF vacuum. Then one naive
question is what is the behavior of the stress tensor in such a
vacuum in the BW. Also, one might expect singularities on the light
cone emanating from the center (the fixed point under the action of
dS group) if we recall that the field perturbations for a massless
scalar in dS grow like $\phisq\sim\chi$, where $\chi$ is the proper
time in dS \cite{vilenkinford,linde,starobinsky,vilenkin} (see also
\cite{hawkingmoss}). The light cone in the RSII model corresponds to
$\chi\to\infty$.

Before we start our discussion, we should mention previous
calculation given in Ref.~\cite{xavi}. In that paper
the stress tensor for a massless minimally coupled scalar
was obtained in four dimensions, in the context of open inflation.
Montes showed that $\tmn$ can be regular everywhere
except on the bubble. As we will see, these
properties hold as well in other dimensions, but only for
massless minimal coupling fields.

\begin{figure}[h]
$$
\begin{array}{ccc}
\includegraphics[width=5cm]{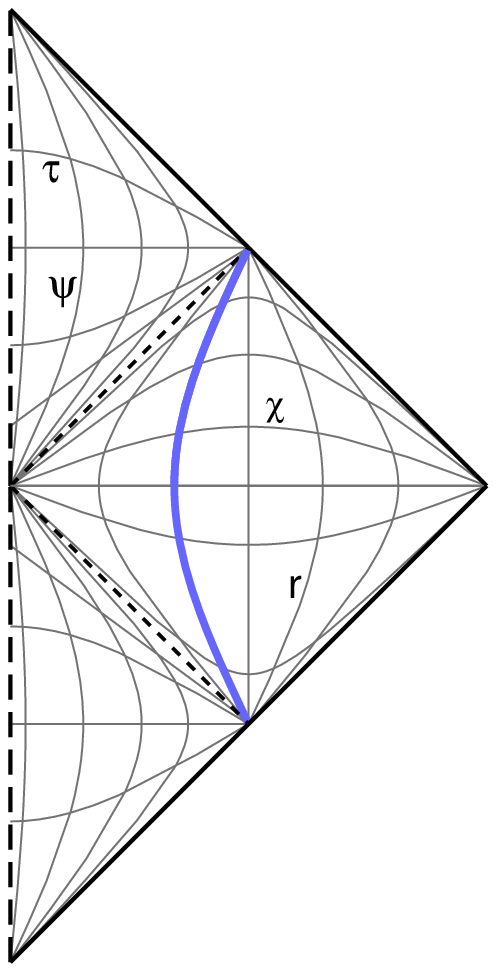}
&\qquad
&\includegraphics[width=3.7cm]{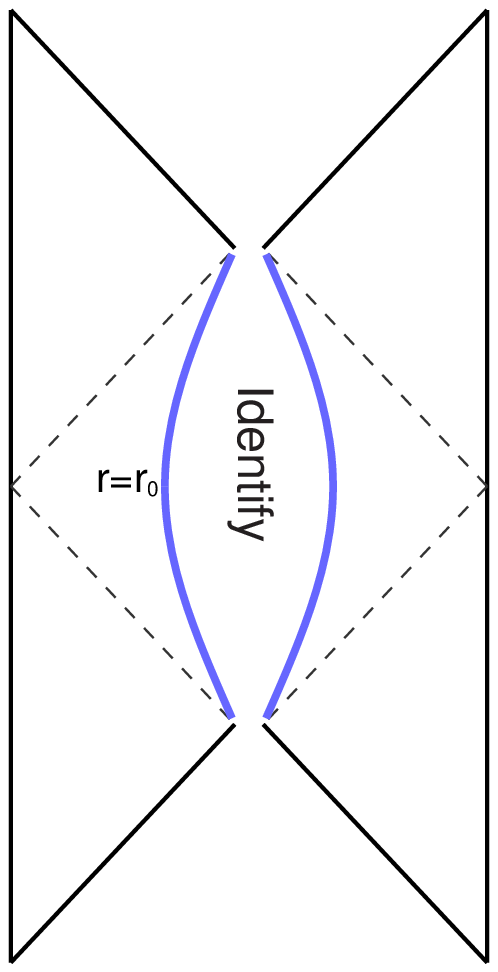} \nonumber\\[-5mm]
{\rm (a)}&\qquad&{\rm (b)}
\end{array}
$$
 \caption{(a) Conformal diagram of the Minkowski space. The surfaces
  corresponding to  constant $r,\chi,\ttau$ and $\psi$ coordinates
  are indicated. The light cone $r=0$ corresponds to the
  dashed lines. (b) Conformal diagram of the VIS model.
  }\label{fig:mink}
\end{figure}

For  simplicity, we consider one extremal case of the RSII model
\cite{rsII} in which
the bulk curvature and hence the bulk cosmological constant
is negligible. We take
into account the gravitational field of the brane by imposing
Israel's matching conditions. The resulting spacetime can be
constructed by the `cut-and-paste' technique. Imposing mirror
symmetry, one cuts the interior of a dS brane in Minkowski and
pastes it to a copy of itself (see Fig.~\ref{fig:mink}). Such a
model was introduced in the context of bubble nucleation by
Vilenkin \cite{v} and by Ipser and Sikivie \cite{is}, and we shall
refer to it as `the VIS model'.

This article is organized as follows. In Section \ref{sec:vis}, we
describe the VIS model and introduce a bulk scalar field with
generic bulk and brane couplings. The Green's function is obtained
first for the case when the bound state is massive, $m_d>0$. The
form of $\tmn$ in the limit $m_d\to0$ is also obtained. In Section
\ref{sec:massless}, we consider an exactly massless bound state
$m_d=0$, and we present the divergences of the AF
vacuum. The case when the bulk mass vanishes is technically
simpler and explicit expressions for $\tmn$ can be obtained. This
is done in Section \ref{sec:M=0}. With this, we describe the
evolution of the field perturbations in Section \ref{sec:pert},
and conclude in Section \ref{sec:concl}.

\section{Scalar fields in the VIS Model}
\label{sec:vis}

In this Section we consider a generic scalar field propagating in
the VIS model,  describing a gravitating
brane in an otherwise flat space \cite{v,is}. Specifically, the
space time consists of two copies of the interior of the brane
glued at the brane location, as illustrated in Fig.~\ref{fig:mink}.

In the usual Minkowski spherical coordinates the metric is
$ds^2=-d\nt^2+d\rrho^2+\rrho^2 d\Omega_{(n)}^2$,  where
$d\Omega_{(n)}^2$ stands for the line element on a unit $n$
sphere. From the symmetry, it is
convenient to introduce another set of coordinates. The Rindler
coordinates, defined by
$$
\begin{array}{ll}
    \rrho= r \cosh \chi~, \\
    \nt= r \sinh \chi~,
\end{array}%
$$
cover the exterior of the light cone emanating from $\rrho=\nt=0$.
In terms of them the brane location is simply $r=r_0$, and the
metric looks like
\begin{equation}
 ds^2=dr^2+r^2 dS_{(\q+1)}^2~,
%     =r_0^2 \, e^{-2\z  }\left(d\z  ^2+ dS_{(\q+1)}^2\right),
\end{equation}
%where here we also introduced the `conformal Rindler' coordinate
%$r=r_0 e^{-\z }$ and $r=r_0$.
where $dS_{(\q+1)}^2$ is the line element of de Sitter (dS) space
of unit curvature radius. Thus, the Hubble constant on the
brane is $H=1/r_0$. In order to cover the interior of the light
cone, we introduce the `Milne' coordinates
according to %$\rrho= \ttau \sinh \psi$ and $t= \ttau \cosh \zeta$,
$$
\begin{array}{ll}
    \rrho= \ttau \sinh \psi~, \\
    \nt= \ttau \cosh \psi ~.
\end{array}%
$$
In these coordinates, the metric is
$ds^2=-d\ttau^2+\ttau^2[d\psi^2+\sinh^2\psi\,d\Omega_{(n)}^2]$.
Note that we can go from the Rindler to the expanding (contracting)
Milne regions making the continuation $r=\pm i \ttau$ and $\chi= \psi
\mp (\pm)i\pi /2$. Here upper and lower signatures correspond to
$+i\epsilon$ and $-i\epsilon$ prescriptions, respectively.

We consider a scalar field even under $Z_2$ symmetry, with generic
couplings described by the action
\begin{equation}\label{action}
    S=-{1\over2}\int_{\rm Bulk}
    \left[(\partial \phi)^2 +
   \left(  M^2 +\xi R\right) \phi^2\right]
    -\int_{\rm Brane}
\left[\mbr+2\xi {\rm tr} \,{\cal K}\right]
        \,\phi^2~,
\end{equation}
where $M$ and $\mu$ are the bulk and brane masses, ${\cal R}$ is
the Ricci scalar and ${\rm tr} \,{\cal K}$ is the trace of the
extrinsic curvature. The latter arises because the curvature
scalar contains $\delta$ function contributions on the brane, and
the factor 2 in front of it is due to the $Z_2$ symmetry. The
stress tensor for a classical field configuration can also be
split into bulk and a surface parts as\cite{bida,saharian}
\begin{eqnarray}\label{tmnclass}
T_{\mu\nu}^{bulk}&=&\left(1-2\xi\right)\,\partial_\mu\phi\partial_\nu\phi-{1-4\xi\over2}\;
\left[%g^{\lambda\sigma}\partial_\lambda\phi\partial_\sigma\phi
\left(\partial\phi\right)^2+\left(M^2+\xi {\cal R}\right)\phi^2
%+{\xi\over n+2}\phi\left\{\Box-M^2\right\}\phi
\right]g_{\mu\nu}
%
%-\xi\phi\big[2\nabla_\mu\nabla_\nu
-2\xi{\cal R}_{\mu\nu}\phi^2
-2\xi\phi\nabla_\mu\nabla_\nu\phi \cr %
T_{ij}^{brane}&=&\left[(4\xi-1)\,\mbr_\eff \,h_{ij}-2\xi\, {\cal
K}_{ij}\right]\,\phi^2\,\delta(r-r_0)~,
\end{eqnarray}
where $h_{ij}$ is the induced metric on the brane, ${\cal
R}_{\mu\nu}$ the Ricci tensor, and the equation of motion has been
used.\footnote{Generically, surface terms are irrelevant for the
Casimir force, but are essential to relate the vacuum energy
density and the Casimir energy, see
\cite{kcd,dc,fulling,saharian,romeosaharian}.}
%
%
%
%
%In the VIS model, the curvature in the bulk is zero. Moreover, the
Here we have introduced an {\em
effective brane mass} as
\begin{equation}\label{meff}
\mbr_\eff\equiv \mbr+2(n+1)\xi H~,
\end{equation}
where $H=1/r_0$ is the Hubble constant on the brane. Then, the
v.e.v. $\tmn$  in point splitting regularization is computed
as\footnote{We omit the anomaly term since it vanishes in the bulk
for odd dimension, and on the brane it can be absorbed in a
renormalization of the brane tension.}
\begin{eqnarray}
\label{splitting}
\la T_{\mu\nu}\ra^{bulk}&=&{1\over2} {\cal
T}_{\mu\nu}
\left[G^{(1)}(x,x')\right]~,%
\end{eqnarray}
with
\begin{eqnarray}
\label{splitting1}
{\cal T}_{\mu\nu} &\equiv&%
\lim_{x'\to
x}\left\{(1-2\xi)\partial_\mu\partial'_\nu-{1-4\xi\over2}\;
g_{\mu\nu}\left(g^{\lambda\sigma}\partial_\lambda\partial'_\sigma+M^2\right)
-2\xi \nabla_\mu\nabla_\nu \right\},
\end{eqnarray}
and
\begin{eqnarray}
\label{splitting2}
\la T_{ij}\ra^{brane}&=&{1\over2}\Big[(4\xi-1)\,\mbr_\eff
\,h_{ij}-2\xi\, {\cal
K}_{ij}\Big]\,\delta(r-r_0)\,G^{(1)}(x,x')\big|_{x'=x}
%-\xi{\cal R}_{\mu\nu}%\cr %
%&-&\delta(r-r_0)\left({1-4\xi\over2}\;\mbr_\eff g_{\mu\nu}+\right)
~,
\end{eqnarray}
where $\partial'_\mu=\partial/\partial x'^\mu$. This expression is
extended to the case with a nonzero bulk cosmological constant by
replacing $M^2$ with $M^2+\xi{\cal R}$ and recovering the Ricci
tensor term in Eq.~(\ref{tmnclass}).

\subsection{Spectrum}
\label{sec:spectrum}

The Klein-Gordon equation following from the action (\ref{action}) is
separable into radial and dS parts so we introduce the mode
decomposition $\phi=\sum\int~\uu_{p}(r)\,{\cal Y}_{p\ell
m}(\chi,\Omega)$, where $m$ is a multiple index. The radial
equation is
\begin{equation}
\label{radial}
 \left[\partial_r^2+{\q+1\over r}\partial_r+{1\over r^2}
   \left(p^2+{\q^2\over 4}\right) -M^2
         \right]\uu_p(r)=0~,
\end{equation}
while the brane terms can be encoded in the boundary condition
\begin{equation}
\label{boundcond} \left.
\left(\partial_r+\mbr_\eff\right)\uu_p\right\vert_{r=r_0}=0,
%\left[-r\partial_r+\mu-{\q\over2}\right]\uu_p\right\vert_{r=r_0}=0,
\end{equation}
where $Z_2$ symmetry has been imposed and the effective brane mass
$\mbr_\eff$ is given in Eq.~(\ref{meff}).

The de Sitter part satisfies
$$
\left[\Box_{n+1} - \left(n/2\right)^2 -p^2 \right] {\cal Y}_{p\ell
m}=0~.
$$
Thus one obtains a tower of modes ${\cal Y}_{p\ell m}$ in dS with
masses $m_\kk^2=(n/2)^2+p^2$ in units of $H$. The mass spectrum
determined by the Schr{\" o}dinger problem defined by
Eqs.~(\ref{radial})
and (\ref{boundcond}). It consists of a bound state plus a continuum
of KK states with $p\geq0$ ($m_{KK}\geq n/2$). The
radial part for the KK modes is of the form
\begin{equation}
\label{kkwavefunct}
 \uu^\kk_p(r)=r^{-\q/2}\left[A_p I_{ip}(Mr) +
               B_{p} I_{-ip}(Mr)\right],
\end{equation}
with $A_p$ and $B_p$ determined by the boundary condition
(\ref{boundcond}) and continuum normalization, %of $\uu_p(r)$,
$2\int dr r^{n-1} \uu^\kk_p(r) \uu^\kk_{p'}(r) = \delta(p-p') $.

\begin{figure}[b]
  \includegraphics[width=8cm]{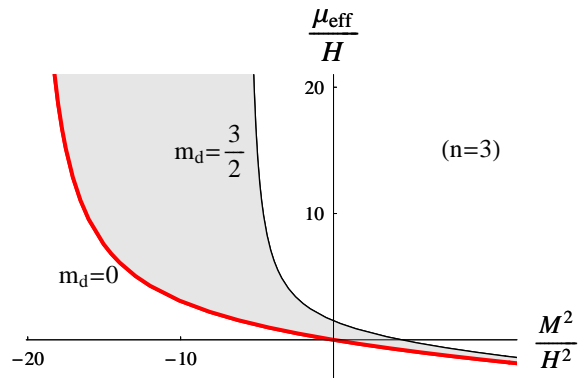}\\
  \caption{
The shaded area corresponds to the values of $\mbr_\eff$ and $M^2$
for which the bound state  is normalizable ($m_d<n/2$) and non
tachyonic, $m_d\geq0$. The thick (red) line corresponds to the
massless case. The plot is for $n=3$. } \label{meffvsM2}
\end{figure}

The mass of the discrete spectrum is
$$
m_d^2=(n/2)^2+p_d^2 < (n/2)^2~,
$$
and hence $p_d$ is pure imaginary. The normalizability implies
that its wave function is
\begin{equation}
\label{wavefunct}
   \uu^\bs(r)=N_d\; r^{-\q/2} I_{-ip_d}(Mr)~,
\end{equation}
with $-ip_d>0$.
%The  normalization constant $N_d$ is given in Appendix
%\ref{sec:cancellation}.
The boundary condition (\ref{boundcond}) determines $p_d$ in terms
of $M$ and $\mbr_\eff$ according to
\begin{equation}
\label{BSmassM}
     \mmu\; I_{-ip_d}(M r_0)- M r_0 I_{-ip_d}'(M r_0)=0~,
%     M r_0 I_{-ip_d}'(M r_0)-\left({n\over2}-\mu_\eff \,r_0 \right)\; I_{-ip_d}(M r_0)=0~,
\end{equation}
where we introduced the combination
\begin{equation}
    \mmu\equiv{n\over2}-{\mbr_\eff\over2H}.%2(n+1)\xi - {r_0 \, \mbr\over2}~,
\label{mu}
\end{equation}
In the limit $M r_0\ll1$ and $\mbr_\eff r_0\ll1$, Eq.~(\ref{BSmassM})
implies that the mass of the bound state is
\begin{equation}
\label{BSmassMsmall} %
%m_d^2= \left(n H-{ \mbr_\eff}\right)\,{\mbr_\eff}
%    + {\mbr_\eff-nH/2\over\mbr_\eff-(n+2)H/2}\; M^2
%    %\left(1-{H\over (1+n/2)H-\mbr_\eff}\right)
%        +{\cal O}\left( M^4\right),
 (H m_d)^2= n H\,{\mbr_\eff}
    + {n\over n+2}\; M^2
    %\left(1-{H\over (1+n/2)H-\mbr_\eff}\right)
        +{\cal O}\left( \mbr_\eff^2, \mbr_\eff M^2, M^4\right),
\end{equation}
which agrees with the results of Ref.~\cite{shs,lasa}.
%\footnote{For $(M r_0)^2\gg1$, it is
%given by
%\begin{equation}
%\label{BSmassMlarge}
%$$
%m_d^2= -2 H M+ {(n+1)(n+3) \over4} H^2 -H \mbr_\eff +{\cal
%O}\left( M^{-1}\right),
%$$
%%\end{equation}
%in this case $H \mbr_\eff$ needs to be large and of opposite sign
%of $M$.}
Figure \ref{meffvsM2} shows the values of $M^2$ and
$\mu_\eff$ for which there exists a non-tachyonic ($-ip_d\leq
n/2$) bound state. In this paper, we are mostly interested in the
situation when the bound state is massless. This happens whenever
Eq.~(\ref{BSmassM}) with $-ip_d$ replaced by $n/2$ holds, that is when
$\mmu$ reaches the `critical' value
\begin{equation}
    \label{muc}
    \mmu_c= {Mr_0 I'_{\q/2}(Mr_0)\over I_{\q/2}(Mr_0)}~.
\end{equation}

\subsection{Green's function}
\label{sec:massiveg}

The renormalized $D$-dimensional Green's function can be split into
the bound state and KK contributions,
\begin{equation}
  G_{(ren)}^{(1)} = G^{\kk}+G^{bs}~, \label{greg}
\end{equation}
with
\begin{eqnarray}
  G^{bs}&=&\uu^\bs(r)\uu^\bs(r')G_{p_d\,(dS)}^{(1)},
\label{greg1}\\
  G^{\kk}&=&\displaystyle\int_0^\infty dp \left[\uu^\kk_p(r )
      \uu^\kk_p(r')\right]^{ren}G_{p\,(dS)}^{(1)}~,
\label{greg2}
\end{eqnarray}
where $G_{p\,(dS)}^{(1)}$ denotes the Green's function of a field
with mass $(n/2)^2+p^2$ in $n+1$ dimensional dS space with $H=1$. It depends
on $x$ and $x'$ through
%the relative `angle' that they subtend, which we call  $\zeta$. %through
the invariant distance in dS, which we call  $\zeta$.
%between the projections of %points $x$ and $x'$
Its precise form is given  in Appendix \ref{sec:GdS}. The
`renormalized' product of the KK mode functions in Eq.~(\ref{greg2}) is
\begin{eqnarray}
\label{uumass} %
\left[\uu^\kk_p(r) \uu^\kk_p(r')\right]^{ren} %
&\equiv& \uu^\kk_p(r ) \uu^\kk_p(r')-\uu^{Mink}_p(r )
\uu^{Mink}_p(r')\cr
    &=& { i\,p \over 2\pi (r r')^{\q/2}}
          {\mmu K_{-ip}(Mr_0) -Mr_0 K'_{-ip}(Mr_0)
                \over \mmu I_{-ip}(Mr_0) -Mr_0 I'_{-ip}(Mr_0)}
                  I_{-ip}(Mr) I_{-ip}(Mr')+
        (p\to -p)~.
\end{eqnarray}
Here, $\uu^{Mink}_p(r )\propto K_{ip}(Mr)$ is the Minkowski
counter part of (\ref{kkwavefunct}). This effectively removes the
UV divergent contribution to the Green's function and guarantees
the renormalized Green's function (\ref{greg})
to be finite in the coincidence limit.

\begin{figure}[tb]
  \includegraphics[width=7cm]{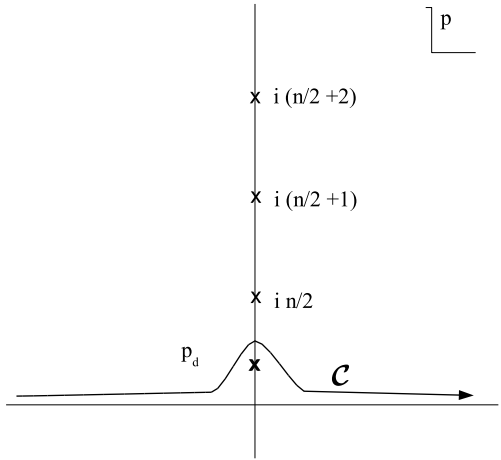}\\
  \caption{Integration contour  in Eq.~(\ref{polesmass}).}\label{contour}
\end{figure}

Since Eq.~(\ref{uumass}) is even in $p$,
Eq.~(\ref{greg2}) can be cast as $G^{\kk}=\int_{-\infty}^{\infty}dp
\,\left[\uu_p^\kk(r)\uu_p^\kk(r')\right]^{ren}_1\big|_{p_d}
\,G_{p\,(dS)}^{(1)}$~, where
$\left[\uu_p^\kk(r)\uu_p^\kk(r')\right]^{ren}_1$ stands for the first term in
Eq.~(\ref{uumass}) only. This can be evaluated summing the residues.
From Eq.~(\ref{gds}), the poles in $G_{p(dS)}^{(1)}$ in the upper $p$
plane are at $p=i(q+ n/2)$, with $q=0,1,2\dots$ (see
Fig.~\ref{contour}). From Eqs.~(\ref{uumass}) and (\ref{BSmassM}), we see
that the KK radial part has a pole at the value of $p$
corresponding to the bound state, $p=p_d$. We shall now show that
the residue is related to the bound state wave function as
\begin{equation}\label{resuu}
2\pi i\; Res \left[\uu_p^\kk(r)\uu_p^\kk(r')\right]^{ren}_1\big|_{p_d}
=-\,\uu^\bs(r) \uu^\bs(r')~.
\end{equation}
Using the Wronskian relation $K_\lambda^{}(z)
I_\lambda'(z)-K_\lambda'(z) I_\lambda^{}(z)=1/z$ %and the relation
%between  $\mmu$ and $p_d$ %
and Eq.~(\ref{BSmassM}), it is straightforward to show that
$$
2\pi i\; Res \left[\uu_p^\kk(r)\uu_p^\kk(r')\right]^{ren}_1
\big|_{p_d}
= -   { p_d / I_{-ip_d}(Mr_0) \over \left.\partial_p
          \left(\mmu I_{-ip}(Mr_0) -M r_0I'_{-ip}(Mr_0)
                \right)\right\vert_{p=p_d}}\;{ I_{-ip_d}(Mr)
        I_{-ip_d}(Mr')\over (r r')^{\q/2}}.
$$
The overall constant in the r.h.s. is nothing but the
normalization constant in the bound state wave function
(\ref{uumass}), up to the sign. Using the Schr\"odinger equation
(\ref{radial}) and integrating by parts, we have
\begin{equation*}
 (p^2-p_d^2) \int_0^{r_0} {dr\over r} I_{-ip_d}(Mr)I_{-ip}(Mr)
   = I_{-ip}(Mr_0) Mr_0  I_{-ip_d}'(Mr_0)
   -I_{-ip_d}(Mr_0) Mr_0  I_{-ip}'(Mr_0).
\end{equation*}
Setting $p=p_d$ after differentiation with respect to $p$, we find
\begin{equation}
\label{radialnorm}
 {1\over N_d^2} = 2 \int_0^{r_0} {dr\over r}
\left[I_{-ip_d}(Mr) \right]^2
   ={1\over p_d}
    I_{-ip_d}(Mr_0) \left. \partial_{p}
      \left(\mmu I_{-ip}(Mr_0)-Mr_0 I'_{-ip}(Mr_0) \right)\right\vert_{p=p_d},
\end{equation}
where we used Eq.~(\ref{BSmassM}). From this and the form of the
bound state wave function (\ref{wavefunct}), it is clear that
Eq.~(\ref{resuu}) holds. Equation (\ref{resuu}) implies that no term
of the form $\uu^{}_\bs(r)\uu^\bs(r') G_{p_d}^{(dS)}$ survives in the
result. This is `fortunate' because close to $r=0$, $\uu^\bs\sim
r^{-n/2-ip_d}$ which is divergent. Thus, Eq.~(\ref{resuu})
guarantees that $\tmn$ is regular on the light cone.

Since only the poles from $G_{p\,(dS)}^{(1)}$ contribute to
$G^{(ren)}$, it can be written as the integral over a contour
${\cal C}$ that runs above $p_d$  (see Fig.~\ref{contour}).
Equation (\ref{gds}) leads to the expression appropriate to see
the coincidence limit,
\begin{eqnarray}
\label{polesmass}
  G^{(ren)}
    &=& \int_{\cal C}dp \,
\left[\uu_p^\kk(r)\uu_p^\kk(r')\right]^{ren}_1
    \,G_{p\,(dS)}^{(1)}\cr
      &=&-  { S_{(\q)}^{-1}
         \over 2^{\q-1}\Gamma\left({\q+1\over 2}\right)}
          \sum_{\s=0}^{\infty}    \sum_{j=0}^{\infty}
          {\mmu K_{\q/2+\s+j}(Mr_0)
          -Mr_0 K'_{\q/2+\s+j}(Mr_0)
                \over \mmu I_{\q/2+\s+j}(Mr_0) -Mr_0 I'_{\q/2+\s+j}(Mr_0)}
\cr  &&\qquad
             {\left({\q\over 2}+\s+j\right)
                 I_{\q/2+\s+j}(Mr)
                 I_{\q/2+\s+j}(Mr')\over (rr')^{\q/2}}
 %  \cr && \qquad     \times
     {(-1)^\s \Gamma\left(\q+2\s+j\right)
     \over j!\,\s !\, \Gamma\left({\q+1\over 2}+\s\right)}
      \left({1-\cos\zeta\over 2}\right)^\s~,
\end{eqnarray}
and %$\nu$ is defined in (\ref{mu}),
we remind that $\zeta$ is the invariant distance in dS.
%and $\left[\uu^\kk_p \uu^\kk_p{}'\right]^{ren}_1$
%is the first term in (\ref{uumass}).
Each term comes from the pole at $p=i((k+j)+n/2)$. Setting $k=0$,
we find that the terms with a large $j$ is unsuppressed for
$r=r'=r_0$ for $\zeta=0$. Hence the Green's function in the
coincidence limit is divergent on the brane. This is the usual UV
`Casimir' divergence near the boundary. Since we are interested in
the IR behavior, we shall not further comment on this UV
divergence.

The term with $\s=j=0$ in Eq.~(\ref{polesmass}) renders the Green's
function  {\em globally} IR divergent in the limit when the bound
state is massless, $\mmu\to\mmu_c$ (see Eq.~(\ref{muc})). One can show
that this term comes from the homogeneous ($\ell=0$) mode of the
bound state. Using Eqs.~(\ref{cancelation}) and (\ref{taylor}), the
leading behavior of Eq.~(\ref{polesmass}) in the massless limit
$m_d\to0$ can be written as
\begin{equation}\label{irdivg}
% this is a G^{(1)}
G^{(ren)(1)}={2\over S_{(n+1)}}\, {H^2\over m_d^2} \,
\uu^\bs_0(r)\uu^\bs_0(r')+{\cal O}(m_d^0)~,
\end{equation}
where $\uu^\bs_0(r)=N_0 I_{n/2}(Mr)/r^{n/2}$ is the wave function
of the bound state (\ref{wavefunct}) for the exactly massless
case.

The divergence (\ref{irdivg}) appears because in the massless
limit, the wave function of the homogeneous mode of the bound
state in the dS invariant vacuum broadens without bound \cite{gaki}. It can
be removed by considering another vacuum with finite width, which
implies breaking dS symmetry \cite{af}. %In order to break dS
%symmetry as little as possible
Later, we shall take the %($O(n+1)$ symmetric)
Allen Follaci vacuum \cite{af}. We will find in Section
\ref{sec:massless} that in the brane world context, this process
removes the global IR divergence, but a localized singularity
within the bulk remains.

Let us now examine the behavior of $\tmn$ in the massless limit,
$m_d\to0$.
% for different
%values of the parameters of the model (\ref{action}).
In this limit, the stress tensor is given by
\begin{eqnarray}\label{IRdivtmn}
\tmn^{bulk}&\simeq &{H^2\over S_{(n+1)}m_d^2} \;{\cal T}_{\mu\nu}
\left[\uu^\bs_0(r) \uu^\bs_0(r')\right]~, \\ %
\label{IRdivtmn2}
\la T_{ij}\ra^{brane}&\simeq& {H^2\over
S_{(n+1)}m_d^2}\, \left[(4\xi-1)\mbr_\eff-2H\xi\right] \;
 ({\uu^\bs_0}|_{r_0})^2 \; \delta(r-r_0)
\,h_{ij}
~,%\Big|_{r'=r}~,
\end{eqnarray}
where ${\cal T}_{\mu\nu}$ is the differential operator given in
Eq.~(\ref{splitting1}) and $h_{ij}$ is the induced metric on the
brane. For $M\neq0$,  $\uu_0^\bs$ is not constant. Therefore
Eq.~(\ref{IRdivtmn}) explicitly shows the presence of a global IR
divergence in $\tmn$ for the dS invariant vacuum in the $m_d\to0$
limit. For $M=0$, $\uu_0^\bs$ is constant. Hence, the bulk part is
finite. However, if $\xi\neq0$ then the surface term
(\ref{IRdivtmn2}) diverges.\footnote{\label{footn:mu} For $M=0$,
$\;\mbr_\eff\,\la\phi^2\ra$ is finite because $\la\phi^2\ra\sim
1/\mbr_\eff$, see (Eq.~\ref{BSmassMsmall}).} Thus, in the limit
$m_d=0$ we are forced to consider another quantum state. This is
the subject of Section \ref{sec:massless}. We shall mention that
there is a possibility to avoid this IR divergence without
modifying the choice of dS invariant vacuum state. In the present
case the divergence is constant and is localized on the brane.
Hence, it can be absorbed by changing the brane tension. We may
therefore have a model in which this singular term is
appropriately renormalized so as not to diverge in the $m_d\to 0$
limit. Of course, such a model is a completely different model
from the original one without this IR renormalization.

The massless minimally coupled limit, $M=\mbr_\eff=\xi=0$, is
exceptional. Both bulk and brane parts of stress tensor
(\ref{IRdivtmn}) and (\ref{IRdivtmn2}) are finite in this limit,
though here there is a slight subtlety. The limiting values depend
on how we fix the ratios among $M, \mbr_\eff$ and $\xi$. For
example, using Eq.~(\ref{BSmassMsmall}), the surface term is given
by
\begin{equation}\label{masminlim}
\lim_{M,\mbr_\eff,\xi\to0} \la T_{ij}\ra^{brane}\simeq
-{n+2\over{}n}{H^2\,({\uu^\bs_0}|_{r_0})^2\over S_{(n+1)}} \;
{\mbr_\eff +2\xi H \over (n+2) H \mbr_\eff + M^2 }
\;\delta(r-r_0)~,
\end{equation}
where we used the approximate mass of the bound state,
$$
m_d^2\simeq n \,\mbr\,H + 2n(n+1)\,\xi\, H^2
    + {n\over n+2}\; M^2 ~,
$$
which is valid in the massless minimal coupling limit
(see Eq.~(\ref{BSmassMsmall})).
Hence, in the absence of any fine tuning
($m_d^2\approx \max(M^2, \xi H^2, H \mbr)$),
it is clear that the
contribution (\ref{IRdivtmn2}) is not large, even though the
Green's function (\ref{irdivg}) is. Thus, only in the case when
the parameters are `fine tuned' according to Eq.~(\ref{muc}) the
stress tensor (\ref{IRdivtmn2}) is large.
From Eq.~(\ref{tmnclass}), if
the Green's function is free from IR divergence,
it is clear that the
brane stress tensor must be zero in the massless minimally coupled
case. The direction that reproduces this result is the one
along which $\la T_{ij}\ra^{brane}$ already vanishes (in the
massive case), that is $\mbr_\eff=-2\xi H$. Note that this feature
is analogous to what happens in dS space \cite{gaki}.

The bulk stress tensor (\ref{IRdivtmn}) also has a similar but
a bit more complicated feature. The operator (\ref{splitting1})
has terms which do not manifestly involve a small quantity, such as
$M^2$, $\xi$ or $\mbr$. However, these terms are associated with
derivative operators. In the limit $M^2\to 0$, we can expand the
$r$-dependence of the term with $k=j=0$ in Eq.~(\ref{polesmass}) as
\begin{equation}
  {I_{n/2}(Mr)\over r^{n/2}}
    \approx {1\over 2^{n/2}\Gamma({n\over 2}+1)}
         \left(1+{M^2 r^2\over 2(n+2)}+\cdots\right).
\end{equation}
Then, the leading term in the above expansion,
which is not suppressed by a factor $M^2$,
vanishes in (\ref{IRdivtmn}). The remaining terms are
finite unless the parameters are fine tuned, as in the case
of the brane stress tensor.

Finally, Eq.~(\ref{IRdivtmn}) also shows that
$\tmn$ is perfectly regular on the light cone for $m_d\neq0$.
As mentioned
before, this happens thanks to the KK modes. Note as well that for
$M\gtrsim H$, Eq.~(\ref{IRdivtmn}) is exponentially localized on the
brane, because the bound state is localized in this case.

\section{Exactly Massless bound state}
\label{sec:massless}

In the preceding section, we have seen that de Sitter (dS) invariant
vacuum causes divergence in
the limit when the bound state is massless.
The divergence is caused by the $\ell=0$ homogeneous mode
in the bound state.
In this section we consider a different choice of
vacuum state for this mode aiming at resolving the problem of divergence,
following the standard methods used in dS space\cite{gaki,af}.
For simplicity, we concentrate on the case of
exactly massless bound state
$m_d=0$ that is $p_d=n i/2$,
although more general cases would be treated in a similar way.
The case with $m_d=0$ includes not only a massless
minimally coupled scalar, $M=\mbr=\xi=0$, but also other fine tuned
cases.

\subsection{The Green's function}
\label{sec:masslessG}

Here, we should split the Green's function into KK and bound
state contributions, $G^{(ren)(1)}=G^{\kk (1)}+G^{bs (1)}$.
We leave the quantum state for the KK contribution untouched,
and change only the contribution from the bound state.
In the integral representation for Green's function in (\ref{polesmass}),
we have used the integration contour
given in Fig.~\ref{contour}.
This choice of contour automatically
takes into account the bound state contribution simultaneously.
In the present case, we consider the contour that runs
below the pole at $p=p_d$
to exclude the bound state contribution.
For $m_d=0$, the integrand has a double pole at $p=ni/2$ because the
pole in the radial modes coincides with one of the poles in the dS
Green's function.
Hence the integral with the contour given in Fig.~\ref{contour},
which runs through these merging poles,
is not well defined.
But the integral with the new contour, which
picks up the contribution from the KK modes only, is well behaved,
and it can be cast as
\begin{eqnarray}
    G^{\kk (1)}&=&2\pi i \Biggl\{
        \sum_{simple~poles} Res\Big(
        \left[\uu_p^\kk(r)\uu_p^\kk(r')\right]^{ren}_1
        G_p^{(dS)(1)}\Big)
        + Res\Big(\left[\uu_p^\kk(r)\uu_p^\kk(r')\right]^{ren}_1
          \Big)_{p_d}\,
        \partial_p\left[ (p-p_d) G_p^{(dS)(1)}\right]\big|_{p_d}
\cr &&\qquad
        + Res\Big( G_p^{(dS)(1)}\Big)_{p_d}
        \partial_p \Big( (p-p_d) \left[\uu_p^\kk(r)\uu_p^\kk(r')\right]^{ren}_1
         \Big)\big|_{p_d}
        \Biggr\}.
\label{gcont}
\end{eqnarray}
As before,
$\left[\uu_p^\kk(r)\uu_p^\kk(r')\right]^{ren}_1$
denotes the first
term in Eq.~(\ref{uumass}).
In the first term the `simple poles' mean poles at $p=i(q+n/2)$ with
$q=1,2,\dots$ (see Fig.~\ref{contour}).
Namely, it is obtained by removing the term with $k=j=0$
from Eq.~(\ref{polesmass}).
The last two terms are contributions from the double pole
at $p=ni/2$.

Next, we consider the contribution from bound state.
A massless bound state behaves as a massless
scalar from the viewpoint of
$n+1$ dimensional dS space. In dS space it is well known that the
dS invariant
Green's function diverges in the massless limit
because of the $\ell=0$ homogeneous mode \cite{af,gaki}. The usual
procedure is to treat separately the $\ell=0$ mode from the rest.
It is easy to show that (see Appendix \ref{sec:GdSmassless})
\begin{eqnarray}
\label{gdisc}
    G^{bs (1)}
%    &=&
%    \uu_d(r) \uu_d(r') \left\{ G_{NZM} +{\cal Y}_{AF}(\chi) {\cal
%    Y}^*_{AF}(\chi')\right\}\cr
    &=& \uu^\bs(r) \uu^\bs(r')\Big\{\sum_{\ell>0} {\cal Y}_{p\ell m}(\chi) {\cal
    Y}^*_{p\ell m}(\chi')+{\widetilde{\cal Y}}_{AF}(\chi) {\widetilde{\cal Y}}^*_{AF}(\chi')\Big\}
     +{\rm c.c.}   \\[1mm]
    &=&\uu^\bs(r) \uu^\bs(r') \left\{  \partial_p\left[ (p-p_d)
            G_p^{(dS)(+)}\right]\big|_{p_d}
        -  \partial_p \left[ (p-p_d) {\cal Y}_{p00}(\chi) {\cal Y}^*_{p00}(\chi') \right] \big|_{p_d}
    + {\widetilde{\cal Y}}_{AF}(\chi) {\widetilde{\cal Y}}^*_{AF}(\chi')
    \right\} +{\rm c.c.}~,\nonumber
\end{eqnarray}
%The first two terms  come from the $\ell\neq0$ modes,
where ${\cal Y}_{p\ell m}(\chi)$ are the positive frequency
dS invariant vacuum modes with  mass $p^2+(n/2)^2$.
%and $G_{NZM}$ stands for the Green's function with the $\ell=0$ mode subtracted
The last term in Eq.~(\ref{gdisc}) is the contribution from the
homogeneous mode in the appropriate state, which we shall take to
be the Allen Follaci (AF) vacuum \cite{af} (see Eq.~(\ref{yaf})).

At this stage, we note that the second term of Eq.~(\ref{gcont})
cancels the first term of Eq.~(\ref{gdisc}), due to Eq.~(\ref{resuu}).
This cancelation resembles the one that occurred in the previous
case between the KK modes and the bound state. In that case, it
guaranteed the absence of light cone divergences. In the present
case, the terms that cancel in Eqs.~(\ref{gcont}) and (\ref{gdisc}) are
already regular. We show below (see discussion after Eq.~(\ref{g0}))
that instead the last term in Eq.~(\ref{gcont}) and the second term
in Eq.~(\ref{gdisc}) diverge on the light cone. However, when added
up, they render $G^{ren}$ finite on the light cone in odd
dimension. (In even dimensions, $G^{ren}$ is finite but its
derivatives are not.)

The fact that the dS invariant Green's function diverges because of the
homogeneous mode implies that
$Res\big(G_p^{(dS)(1)}\big)_{p_d}=\lim_{p \to p_d } (p-p_d){\cal
Y}_{p00} {\cal Y}'^*_{p00}+{\rm c.c.}$. Using this and
Eq.~(\ref{resuu}), we can rewrite the total Green's function in the
more convenient form
\begin{equation}
\label{GmasslesBS}
        G^{(ren)(1)} = G_{simple}^{(1)}  + G_{LC}    +
        {\widetilde G}_{AF} ~,
\end{equation}
with
\begin{eqnarray}
\label{GmasslesBS2}
    G_{LC} &\equiv& 2\pi i \, \partial_p\Big(
        (p-p_d) {\cal Y}_{p00} {\cal Y}'^*_{p00}
            \;\;(p-p_d) \left[\uu_p^\kk(r)\uu_p^\kk(r')\right]^{ren}_1
                \Big) \big|_{p_d}~+{\rm c.c.}~, \cr
    {\widetilde G}_{AF}  &\equiv& \uu^\bs(r) {\uu^\bs}(r') \,{\widetilde {\cal Y}}_{AF}
    {{\widetilde {\cal Y}}'^*}_{AF} ~+{\rm c.c.}~,
\end{eqnarray}
and $G_{simple}^{(1)}$
is the contribution from the first term in Eq.~(\ref{gcont}).
As is manifest
from the expression (\ref{polesmass}) with the $k=j=0$ term removed,
$G_{simple}^{(1)}$
is regular in the coincidence limit, everywhere except for
the `Casimir' divergences on the
brane, and depends on the points $x$ and $x'$ through $r$, $r'$
and $\zeta(x,x')$ (the invariant distance in dS between the
projections of points $x$ and $x'$), and hence it is dS invariant.
%${\chi,\Omega}$ and ${\chi',\Omega'}$).
The second term, $G_{LC}$, contains the two contributions in
Eqs.~(\ref{gcont}) and (\ref{gdisc})
%from the bound state and from the KK modes
that are separately divergent on the light cone mentioned in the
previous paragraph. %We shall see that in odd dimension they cancel
%and that they do only partially in even dimension.
The last term, ${\widetilde G}_{AF}$, encodes the choice of vacuum
for the zero mode of the bound state.

~\\

Let us examine the contribution potentially divergent on the light
cone $G_{LC}$. The derivative of the $\chi$ dependent part is
obtained from Eqs.~(\ref{cpsmall}) and (\ref{yp00}). The derivative of
the radial part follows from Eq.~(\ref{uumass}) and
$\partial_\lambda I_\lambda(z)= I_\lambda(z)\log z+f(z)$, where
$f(z)$ is a regular function. It is easy to see that $G_{LC}$
takes the form
\begin{eqnarray}\label{g0}
G_{LC}^{}(x,x')&=& {2\over nS_{(n+1)}}\;\uu^\bs(r)\uu^\bs(r')
\;{\rm Re}\,\big[ F(x)+F(x') \big] + {\rm regular},
\end{eqnarray}
with
\begin{eqnarray}\label{g02}
F(x)&=&\log r+n\int_0^\chi {d\chi_{{}_1}\over \cosh^\q
\chi_{{}_1}}
    \int_{-{i\pi\over 2}}^{\chi_{{}_1}}
            d\chi_{{}_2} \cosh^{\q} \chi_{{}_2} ~,
\end{eqnarray}
where the indicated regular term depends on $r$ and $r'$ only.
% and vanishes for $M=0$.
The double integral in (\ref{g02}) grows linearly with $\chi$
for large $\chi$ and eventually blows up on the light cone (The
first integral asymptotically grows like $e^{n\chi_1}/n$, and therefore the
integrand in the second integral goes to a constant). This is the
expected behavior from the massless bound state. On the other
hand, the KK modes contribute the $\log r$ term, which cancels the
light cone divergence. To see this, we integrate by parts to obtain
\begin{equation}\label{doubleint}
F(x)=\log r+ \log\sinh\chi+ \int_0^\chi {d\chi_{{}_1}\over
\cosh^\q \chi_{{}_1}}
    \int_{-{i\pi\over2}}^{\chi_{{}_1}}
            d\chi_{{}_2}
            {\cosh^{\q}\chi_{{}_2}\over\sinh^2\chi_{{}_2}}~,
\end{equation}
and now the double integral is bounded. The first two terms are
simply  $\log\,\nt$. Thus,
the leading divergence in $G_{LC}$ on the light cone
cancels between contributions from bound state and
from the KK modes,
although it still diverges logarithmically at infinity.

This statement has to be qualified for even dimension. In this
case, the derivatives of $G_{LC}$ diverge on the light cone
because of the last term in Eq.~(\ref{doubleint}). To see this, note
that the integrand in Eq.~(\ref{doubleint}) can be expanded in
exponentials. Then, the integral is a sum of exponential terms
except for one, of the form $\chi e^{-n \chi}$, if $n$ is even.
In terms of the null coordinates $U$ %=\rrho+\nt=re^\chi$
and
$V$, %=\rrho-\nt=re^{-\chi}$
this is $\sim \left({V/U}\right)^{n/2} \,\log\left({V/ U}\right)$
for $\chi\to+\infty$ (for $\chi\to-\infty$ replace
$U\leftrightarrow V$). Even though the Green's function is regular
at $V=0$, the stress tensor develops a singularity which behaves like
$\sim 1/V$ or $1/U$ on the light cone in four dimensions $(n=2)$
and like $\sim \log V$ or $\log U$ in six dimensions $(n=4)$ if
$\xi\neq0$.

{}For reference, we show the explicit form of $F(x)$ for
dimensions 4 and 5,
\begin{equation}\label{F234}
%\int_0^\chi {d\tilde\chi\over \cosh^\q \tilde\chi}
%         \int_{-\pi i\over 2}^{\tilde\chi}
%            d\hat\chi \cosh^{\q} \hat\chi=
F(x)=
\left\{%
\begin{array}{ll}
\displaystyle
    {U \log U + V\log |V| \over \rrho} %+ ~(x\to x')
        & \hbox{for}\qquad n=2 \ ,\\[4mm]
\displaystyle \log\left(\nt+\ttau\right)-{(\nt-\ttau)\ttau\over\rrho^2} %+ ~(x\to x')
 & \hbox{for}\qquad n=3 \ ,
 %\\[4mm]
%\displaystyle
%    {(2\rrho-t) u^2 \log u + (2\rrho+t) v^2\log |v| \over 4\rrho^3}
%        + {T^2\over 2\rrho^2} %+~ (x\to x')
%            \qquad
%    & \hbox{for}\qquad n=4 \ ,\\
\end{array}%
\right.
\end{equation}
where $\ttau^2={\nt}^2-\rrho^2$. Note that despite appearances,
Eq.~(\ref{F234}) is regular at $\rrho=0$. This is guaranteed since the
$\chi$ dependent part of $F(x)$ is related to the
dS invariant vacuum
modes (see Eq.~(\ref{yp00})), which are regular at
$R=0$ ($\xi=-\pi i/2$) by construction.
%from the lower index in the first integral in (\ref{g0}).
The expressions (\ref{F234}) are appropriate in the Milne region, and we have
already taken the real part, which is the relevant part for $G^{(1)}$.

~\\

The last term $\tilde G_{AF}$
in Eq.~(\ref{GmasslesBS}) corresponds to the choice
of vacuum for the $\ell=0$ mode. This mode is peculiar because it
behaves like a free particle rather than an oscillator. The
eigenstates of the Hamiltonian, and in particular the ground
state, are plane waves in field space. However, such states are
not normalizable. One can construct well defined states as wave
packets. The simplest option is a Gaussian packet. This is the
Allen Follaci vacuum \cite{af}, which in fact is a two-parameter
family of vacua.  Its mode function is given by \cite{af}
\begin{equation}
\label{yaf}  %
{\widetilde {\cal Y}}_{AF}(\chi)={1\over
\sqrt{S_{(n)}}}\left[{1\over 2\alpha} + i \beta -i\alpha
\int_0^\chi {d\chi'\over \cosh^\q \chi'}\right]~,
\end{equation}
where $\alpha>0$ and $\beta$ are the mentioned free (real)
parameters. We shall impose that the vacuum is time reversal
symmetric. This translates into ${\widetilde {\cal
Y}}_{AF}(-\chi)={\widetilde {\cal Y}}_{AF}^*(\chi)$, and implies
$\beta=0$. Because of the time dependence, it breaks dS symmetry.
For this vacuum,
\begin{equation}\label{gaf}
%this is a G^{(1)}
{\widetilde G}_{AF}(x,x')={\uu^\bs(r) \uu^\bs(r')\over S_{(n)}}
\left\{ {1\over 2\alpha^2}+2\alpha^2 \textrm{Re}\left[F_{AF}(x)
F_{AF}^*(x')\right] - \textrm{Im}\left[F_{AF}(x)+
F_{AF}(x')\right] \right\}
\end{equation}
with
\begin{eqnarray*}
F_{AF}=\int d\chi\,\cosh^{-n}\chi.
\end{eqnarray*}
In order to obtain the
analytic continuation of $F_{AF}$ to the Milne region, it is
better to write it as an integral over a contour with constant
$T$. It is straightforward to see that
$$
F_{AF}(x)=-T\int^{\rrho} {d\rrho' \over \rrho'^n} %
\left(\rrho'^2-T^2\right)^{{n\over2}-1}~.
$$
It is transparent now that $F_{AF}$ behaves like $1/\rrho^{n-1}$
near $\rrho=0$. It is also clear that it is regular on the light
cone and is bounded at infinity. Moreover, it gets an imaginary
part in the Milne region for odd dimensions (in the Rindler region
it is always real). More specifically, we have
\begin{equation}\label{FAF234}
F_{AF}(x)=
\left\{%
\begin{array}{ll}
  \displaystyle
    {\nt\over\rrho}~, \qquad
  & \hbox{for}\qquad n=2, \\[2mm]
  \displaystyle
  {i\over2} \left[ {\nt\,\ttau\over\rrho^2}  -\ln\left( i {\nt+\ttau\over\rrho} \right)  \right]  \qquad
  &\hbox{for}\qquad n=3~.%\\
%  \displaystyle
%    {t\over\rrho}-{t^3\over3\rrho^3}~, \qquad
%  & \hbox{for}\qquad n=4
\end{array}%
\right.
\end{equation}

\subsection{Divergent stress tensor}
\label{sec:masslesstens}

Now we discuss the form of the expectation value of the stress
tensor for the possible values of $M$, $\mbr_\eff$ and $\xi$ in
which the bound state is massless.
%From the form of the stress
%tensor (\ref{splitting}), it follows that there are essentially
%There are basically three types of situations:
%corresponding to the presence of $\phisq$ terms in $\tmn$
%the massless minimally coupled case, the
%$M=0$ case (with $\xi\neq0$), and the case with $M\neq0$.\\
The bulk part of the stress tensor is most conveniently separated
into
\begin{eqnarray}
\tmn&=&\tmn_0+\tmn_{simple}
\end{eqnarray}
where $\tmn_0$ contains the contributions from $G_{LC}$ and
${\widetilde G}_{AF}$, whereas $\tmn_{simple}$ is the
contribution from $G_{simple}^{(1)}$. All the IR irregularities are
contained in $\tmn^{bulk}_0$. From Eqs.~(\ref{g0}) and
(\ref{gaf}), we obtain the bulk part of $\tmn_0$ as
\begin{eqnarray}\label{t0}
\tmn^{bulk}_0 & = & %
%{\uu^\bs^2 \over S_{(n)}} \;
{1 \over 2S_{(n)}} \; {\cal{}T}_{\mu\nu}\,\Big[\uu_{bs}(r)
\uu_{bs}(r')
    \left\{{1\over 2\alpha^2}+
      2\alpha^2 \textrm{Re}\big[F_{AF}(x)F_{AF}^*(x')\big]\Big\} %
      \right]\cr
&& -2\xi\;%
%\uu^\bs^2%
\;\nabla_\mu\nabla_\nu \;\uu_{bs}(r) \uu_{bs}(r')
   \left[ {{\rm Re}\, F(x)\over n S_{(n+1)}}-
{\textrm{Im} \,F_{AF}(x) \over 2 S_{(n)}} \right]~.
\end{eqnarray}
The term proportional to $\alpha^2$
diverges at $R=0$ like $1/R^{2n}$ for any value of
$\xi$. $R=0$ corresponds to a timelike
axis in the bulk passing through the center of symmetry
(see Fig.~\ref{diagram}).
The last term also diverges like $1/R^{n+1}$ for odd dimensions,
but vanishes for even dimensions.
The piece involving ${\rm Re}\,F$ diverges on the light
cone for $n=2$ or 4.

Let us begin with the most general case with $M\neq0$.
In this case $\uu^\bs(r)$ is not
constant, and the contribution from the
term inversely proportional to $\alpha^2$ in Eq.~(\ref{t0})
does not vanish.
%\begin{equation}\label{tmnM}
%\tmn^{bulk} \supset {1\over S_{(n)}}\;\left\{
%{1\over4\alpha^2}\;{\cal T}_{\mu\nu} \left[\uu^\bs(r)\uu^\bs(r')\right]
%+\alpha^2\;{\cal T}_{\mu\nu} \left[\uu^\bs(r)\uu^\bs(r'){\rm
%Re}(F_{AF}F_{AF}'^*)\right]\right\} ~.
%\end{equation}
This term is analogous to Eq.~(\ref{IRdivtmn}), and it diverges
globally in the  $\alpha\to 0$ limit.
Hence, this state cannot be taken, and one has to content
with the AF vacuum with some nonzero $\alpha$. Hence, the
term proportional to $\alpha^2$
is unavoidable. But this is very noticeable since it contains
a singularity at $R=0$ of the form $\sim \alpha^2/\rrho^{2n}$,
present even for minimal coupling.
% and in particular breaks dS invariance.
%The factorized terms in the ${\widetilde G}_{AF}$ also contribute weaker
%divergences at $R=0$ of the type $1/R^{n}$ for $\xi=0$ and
%$1/R^{n+1}$ for $\xi\neq0$ in odd dimensions. Also, in four
%dimensions, the terms $\uu^\bs(r) \uu^\bs(r') (F+F')$ from $G_{LC}$ give
%a $1/U$ divergence on the light cone as mentioned before.
The main point is that in the presence of a bulk mass, $\tmn$
contains a quite severe bulk singularity even after we get rid of
the global IR divergence. There is of course the possibility that
a different choice of vacua for the KK modes could cancel it out.
In this case, it seems that the vacuum choice should not be dS
invariant, otherwise the singular zero mode contribution that is not
dS invariant could not be
compensated. We leave this issue for future research.

%In the case of $\xi\neq0$ with $M=0$ we obtain basically
%the same conclusion: the dS
%invariant vacuum is not allowed and the AF vacuum induces a
%singularity in the bulk.
%Only for massless minimal coupling a dS
%invariant state (the Garriga Kirsten vacuum) generates a stress
%tensor free from IR divergences.

Now we turn to the case with $M=0$. Let us first consider the
nonminimal case $\xi\neq0$. Since $M=0$, $\uu^\bs(r)$ is constant
and the bulk stress tensor does not diverge globally in the
$\alpha\to 0$ limit. Divergences in the other $\alpha$ independent two
terms in (\ref{t0}) also vanish in even dimensions with $n\geq
6$. However, we cannot avoid the divergence in the brane stress
tensor in the $\alpha\to 0$ limit. From Eq.~(\ref{splitting2}) and
taking into account that $\mbr_\eff=0$ (see
Eq.~(\ref{BSmassMsmall})), we have
\begin{eqnarray}\label{tbrane}
\la T_{ij}\ra_0^{brane} &=&%
-2\xi \la\phi^2\ra_{AF}\;{\cal K}_{ij}\;\delta(r-r_0)\nonumber\\[2mm]%
&=& -{\xi\,n\over r_0^{n+1}} \left({(2\alpha)^{-2}+\alpha^2
F_{AF}^2(x)\over S_{(n)}}+ {2{\rm Re}\,F(x)\over S_{(n+1)}}
\right)\,h_{ij}\;\delta(r-r_0) ~,
\end{eqnarray}
where we used that $\uu_\bs^2=n/2r_0^n$.
Thus, one is forced to take the AF vacuum
again. As a result,
% $\tmn^{brane}$ depends on time, but most notably,
the bulk stress tensor develops the same singularity at
$R=0$ as before in the $M\ne 0$ case.
The globally divergent ($1/\alpha^2$) term in
Eq.~(\ref{tbrane}) is proportional to the induced metric. One might
wonder whether this effect is physical or not, since it could be
simply absorbed in the brane tension as before.
We think that such a procedure
is not justified here because $\alpha$ is a state dependent parameter, and
renormalization should be done independently of the choice of
the quantum state.

%minimal
Before we examine the stress tensor in the massless minimal coupling
case, let us comment on the relation between the AF vacuum and the
Garriga Kirsten (GK) vacuum.
The latter was introduced in \cite{gaki} for a massless minimally
coupled scalar in dS. It corresponds to the plane wave state with
zero momentum in field space. This is intrinsically ill defined,
giving a `constant infinite' contribution to the Green's function.
However, the point is that, since it is an eigenstate of the
Hamiltonian, it does not depend on $\chi$ so it is dS invariant.
Divergence in the Green's function
can be accepted in the massless minimally coupled case.
Our reasoning is as follows. In this case,
the action has the symmetry $\phi\to\phi+$ constant.
If we consider it as a `gauge' symmetry,
all the observables are to be constructed from derivatives
(or differences) of the field.
In fact the stress tensor operator (\ref{splitting1}) only
contains derivatives of the type $\partial_\mu\partial'_\nu
G^{(1)}(x,x')$ in this case.
Then we will find that the constant contributions in the Green's
function are
irrelevant.\footnote{Taking this vacuum is analogous to performing
a Gupta-Bleuler quantization \cite{renaud}.}
We need a practical way to compute quantities in this vacuum.
If we follow the argument by Kirsten and Garriga, it will be given
by the limit of AF vacuum
$$
|0\ra_{GK}\equiv\lim_{\alpha\to0} |0\ra_{AF}~.
$$
{}From Eq.~(\ref{tbrane}), it follows that the
brane term $\la T_{ij}\ra^{brane}$ vanish in the GK vacuum.
The first term% inversely proportional to $\alpha^2$
in Eq.~(\ref{t0}), which is responsible for the
`infinite constant', vanishes for $M=0$.
While the second term proportional to $\alpha^2$
also vanishes in the $\alpha\to 0$ limit.
We shall note that the last two terms in Eq.~(\ref{t0})
independent of $\alpha$ are also zero for $\xi=0$.
Hence, we are left with $\tmn_{simple}$, which is manifestly
dS invariant, and
is finite (aside from the `Casimir' divergences
on the brane).
Thus, the total stress tensor in the GK vacuum is given by
a simple formula presented below in
Eq.~(\ref{tmnM=m=0}) with $\xi=0$.
In contrast to the massless minimal coupling limit
discussed in the preceding section, here we do not have
any ambiguity.

\section{Zero bulk mass}
\label{sec:M=0}
In the absence of bulk mass $M$, the Green's function and the
stress tensor can be obtained explicitly. We shall thus discuss now
this case.

\subsection{Generic mass of the bound state}

First we discuss the case when the bound state is not massless.
The other case
is postponed until Section \ref{sec:M=m=0}.

For the bound state, the boundary condition (\ref{boundcond})
fixes $p_d=i\mmu$  and $m_d$ is given by Eq.~(\ref{BSmassMsmall}) with
$M=0$. Its wave function is proportional to
$r^{-ip_d-n/2}=r^{-\mbr_\eff/H}$, which is constant if $m_d=0$
(see Eq.~(\ref{BSmassMsmall})). From the discussion in Section
\ref{sec:spectrum}, the bound state is normalizable and
non-tachyonic for $n/2\geq\mmu> 0$ ({\em i.e.} $0\leq
\mbr_\eff<(n/2)H$), see Fig.~\ref{meffvsM2}.

The normalized radial KK modes are
\begin{equation}
 \uu^\kk_p(r)=\sqrt{2\over \pi r^{\q}
   (1+(\mmu/p)^2)}\left(
   \cos \left( p\z  \right)
         +{\mmu\over p}\sin \left( p\z   \right)\right)~,
\end{equation}
and the renormalized product of mode functions  in Eq.~(\ref{greg2}) is
\begin{equation}
\label{uumassless}
\left[\uu_p^\kk(r)\uu_p^\kk(r')\right]^{ren}
   ={1\over 2\pi r^{{\q/ 2}}}
    \left[{p+i\mmu\over p-i\mmu}\; (rr')^{-ip}
          +{p-i\mmu\over p+i\mmu} \;(rr')^{ip}\right]~.
\end{equation}

Proceeding as in Eq.~(\ref{polesmass}), we can explicitly perform the
integration over the KK modes and in this case, we
obtain a simpler expression,
\begin{equation}
\label{polesmassless}
  D_{(ren)}^{(1)}(x,x')
   =  {1 \over  (2r_0)^{\q} S_{(\q)}\Gamma\left({\q+1\over 2}\right)}
          \sum_{\s=0}^{\infty}    \sum_{j=0}^{\infty}
          {{\q\over 2}+\mmu+\s+j \over {\q\over 2}-\mmu+\s+j}
          ~{(-1)^\s \Gamma\left(\q+2\s+j\right)
            \over j!\,\s !\, \Gamma\left({\q+1\over 2}+\s\right)}
          \left({rr'\over r_0^2}\right)^{\s+j}
          \left({1-\cos\zeta\over 2}\right)^\s~,
\end{equation}
where as usual, we denote the (bulk-) massless Green's function by
$D_{(ren)}^{(1)}(x,x')$. From Eq.~(\ref{polesmassless}), it is
manifest that the Green's function in the dS invariant vacuum state is
regular at $r=0$ in the coincidence limit. As before, in the
massless limit %$m_d\to0$
$\mmu\to \q/2$, it is singular. However, since
the divergent term $k=j=0$ is constant in the present case,
it will not affect the bulk $\tmn$.

In order to compute the stress tensor, it is convenient to rewrite
Eq.~(\ref{polesmassless}) in a more compact form. For a conformally
coupled field, $\mmu$ vanishes, and the Green's function actually takes the
simple form\footnote{From Eq.~(\ref{polesmassless}), the expression
for a Dirichlet scalar ($\mmu\to\infty$) is the same with opposite
sign.}
\begin{equation}
\label{mu=0}
  D^{(ren)(1)}_{\mmu=0}(x,x') =
{1\over n S_{(\q +1)}}  \left({1\over r_0^2 + (r r'/r_0)^2 - 2 r
r' \cos\zeta} \right)^{\q /2}.
\end{equation}
This expression can be obtained by the method of images. It
corresponds to the potential induced by a source of unit charge at
$x'$ together with an image source located at $r'_I=r_0^2/r'$ with
a charge $q'_I=(r'_I/r_0)^n$. From the form of
(\ref{polesmassless}), the Green's function for $\mmu\neq0$ can be
obtained from Eq.~(\ref{mu=0}) by applying the integral operator
\begin{equation}
\label{integral}
      D^{(ren)(1)} =
      \left[1 + 4\mmu r_0^{-2\mmu}
          \int_{r_0}^\infty {d \tilde r_0\over \tilde r_0} \tilde r_0^{2\mmu} \right]
      D^{(ren)(1)}_{\mmu=0}~.
\end{equation}

Borrowing the intuition from the method of images,
Eq.~(\ref{integral}) can be interpreted as the potential induced by
the image charge mentioned in the previous paragraph together with
a string stretching from $r_0^2/r'$ to infinity %$r\to\infty$
along the radial direction defined by $x'$ with a charge line
density given by $\lambda(r)=4\mmu (r/r_0)^{2\mmu-1}/r'$. To
obtain the stress tensor, we can first compute the case $\mmu=0$
and then apply the same operator as in Eq.~(\ref{integral}). The
general result with  $M=0$ and  $m_d\neq0$ is
\begin{eqnarray}
    \label{tmassless}
    {}^{(m_d\neq0)}\la T^r_{~r}\ra =  (\xi-\xi_c) { (-1)^n (\q +1)\over S_{(\q+1)} r_0^{\q +2}}
    &\Biggl\{&
        {r_0^{2\q +2} \over (r_0^2-r^2)^{\q +1} }
        + {4\mmu\over \q +2-2\mmu} ~
        {}_2F_1\left(\q +1, {\q\over2} +1-\mmu;{\q\over2} +2-\mmu;{r^2\over r_0^2}\right)
    ~\Biggr\}~, \\
    &~&\cr
    {}^{(m_d\neq0)}\la T^i_{~j}\ra =  (\xi-\xi_c){ (-1)^n (\q +1)\over S_{(\q+1)}
        r_0^{\q +2} } &\Biggl\{&
    {r_0^{2\q+2}  (r_0^2+r^2) \over (r_0^2-r^2)^{\q +2}}
    + {4\mmu\over \q +2-2\mmu} ~
        {}_2F_1\left(\q +1, {\q\over2} +1-\mmu;{\q\over2} +2-\mmu;{r^2\over r_0^2}\right)
        \cr
%    &&\qquad\qquad\qquad\qquad\qquad
    &&\qquad\qquad\qquad {8\mmu\over \q +4-2\mmu} {r^2\over r_0^2} ~
    {}_2F_1\left(\q +2,{\q\over2} +2-\mmu;{\q\over2} +3-\mmu;{r^2\over r_0^2}\right)
    ~\Biggr\}\delta^i_{~j}~,\nonumber
\end{eqnarray}
where $\xi_c=n/4(n+1)$. As an aside, we shall note as well that
for conformal coupling, $\tmn^{bulk}=0$ even with a nonzero
boundary mass $\mbr$. This is a consequence of conservation and
tracelessness of $T_{\mu\nu}$.

\subsection{Massless bound state}
\label{sec:M=m=0}

We consider now the case $m_d=0$, that is $\mmu=n/2$. We can
proceed as in Eq.~(\ref{GmasslesBS}) and decompose
$D^{ren}=D_{simple}+G_{LC}+{\widetilde G_{AF}}$, where
$D_{simple}$ is the contribution from the simple poles
$p=i(n/2+k)$ with $k=1,2,\dots$ in Eq.~(\ref{gcont}). Thus, it is
given by the terms in Eq.~(\ref{polesmassless}) with non-vanishing $k$
and $j$. The integral representation analogous to Eq.~(\ref{integral})
is now
\begin{equation}\label{dintegral}
   D_{simple}^{(1)}
   =\left[1 + {2 n \over r_0^n}
        \int_{r_0}^\infty {d \tilde r_0\over \tilde r_0} \tilde r_0^n \right]
    \left( D^{(ren)}_{\mmu=0} - {1\over n S_{(n+1)} \tilde r_0^n}
    \right)~,
\end{equation}
where we subtract the constant to remove the $j=k=0$ term. The
explicit expression in four dimensions was obtained in
\cite{xavi}\footnote{In Ref. $Z_2$ reflection symmetry is not
assumed for the field.}, and we shall not reproduce it here. The
case of main interest for us is $n=3$ and we find, up to an finite
constant,
\begin{eqnarray}\label{gsimple}
   D_{simple}^{(1)}(x,x')
={\rm Re}\,\Biggl[{1\over 8 \pi^2} {1\over r_0^3\Delta^{3/2}}
    +{3\over 8\pi^2 r_0^3} \left\{
    {\cos2\zeta\over\sin^2\zeta}-{\cos2\zeta -(rr'/r_0^2)\cos\zeta
    \over \sin^2\zeta \;\Delta^{1/2}}
    -\log\left[1 -{rr'\over r_0^2}\cos\zeta +\Delta^{1/2} \right]
\right\}\Biggr]~,
\end{eqnarray}
where $\Delta=1+(rr'/r_0^2)^2-2 (rr'/r_0^2) \cos\zeta $. As
mentioned above, in the coincidence limit this contribution is
regular except on the brane. We shall note as well that it grows
logarithmically at infinity.

~\\

The contribution $\tmn^{bulk}_{simple}$ is easily found exploiting
the integral representation (\ref{dintegral}), as before. The only
difference between (\ref{dintegral}) and (\ref{integral}) in the
limit $\mmu=n/2$ ({\em i.e.} $m_d=0$) is a constant, which does
not affect $\tmn$ in this case. Thus, $\tmn^{bulk}_{simple}$
reduces to (\ref{tmassless}) with $\mmu=n/2$. This gives a simple
expression in terms of elementary functions. In four dimensions it
is given in \cite{xavi} (for $\xi=0$). In five dimensions, we
obtain
\begin{eqnarray}
    \label{tmnM=m=0}
    \la T^r_{~r}\ra^{bulk}_{simple}  &=& -{9\over32\pi^2}\;{\xi-\xi_c\over r_0^3}\;\left\{
        {r_0^{6}\over (r_0^2-r^2)^{4} } +{r^4-3r_0^2 r^2+3r_0^4
        \over (r_0^2-r^2)^{3} } \right\}~,\nonumber\\[2mm]
    \la T^i_{~j}\ra^{bulk}_{simple}  &=& -{9\over32\pi^2}\;{\xi-\xi_c\over r_0^3}\;\left\{
        r_0^{6}{r^2+r_0^2 \over (r_0^2-r^2)^{5} }
    +{r^2\over2} {r^4-4r_0^2 r^2+6r_0^4
        \over (r_0^2-r^2)^{4} }
        +{r^4-3r_0^2 r^2+3r_0^4 \over (r_0^2-r^2)^{3} }
     \right\}\delta^i_{~j}~,
\end{eqnarray}
which is dS invariant and regular everywhere except on the brane.

\section{Perturbations}
\label{sec:pert}

We shall now discuss the form of the field perturbations,
focusing on the bulk massless case with $n=3$, since we have obtained a
closed form expression for the Green's function and
the stress tensor in the preceding section. We begin
by the case with generic coupling, and we consider
$\la\phi^2(x)\ra$ in the Allen Follaci vacuum. The massless
minimally coupled case in the de Sitter (dS) invariant Garriga Kirsten vacuum
requires a different treatment, which will be discussed in
Section~\ref{sub:mmc}.

\subsection{Generic coupling}

The renormalized expectation value of $\phi^2(x)$ is given by
$$
\la\phi^2(x)\ra^{ren}_{AF}={1\over2}\,G_{ren}^{(1)}(x,x)~,
$$
with $G_{ren}^{(1)}$ given by (\ref{GmasslesBS}). From Eqs.~(\ref{gaf})
and (\ref{FAF234}), it diverges at $\rrho=0$ because of $G_{AF}$.
It is also clear that this contribution is
bounded at (null) infinity. From Eq.~(\ref{g0}), the $G_{LC}$
contribution is regular on the light cone. Equation
(\ref{gsimple}) shows that the KK contribution $D_{simple}$
diverges on the brane. In the bulk, both $D_{simple}$ and $G_{LC}$
grow logarithmically at infinity. However, the growing terms
cancel out. %$$\la\phi^2(x)\ra^{ren}_{AF}$$
Indeed, Eq.~(\ref{gsimple}) shows that as long as $x$ is not on the
brane, $D_{simple}$ at coincident points is dominated by the
logarithmic term. Therefore we have
\begin{equation}\label{gsimpleinf}
{1\over2}D_{simple}^{(1)}(x,x)\sim -{3\over 16\pi^2r_0^3}\,\log
\Big|1-{r^2\over r_0^2}\Big|~,
\end{equation}
in the limit $x\to\infty$. As for $G_{LC}$, since the wave
function of the bound state in the $M=0$ case is
$\uu_\bs^2=n/2r_0^n$, we find
\begin{equation}\label{g0inf}
{1\over2}G_{LC}^{}(x,x)={3\over 8\pi^2r_0^3}\,{\rm
Re}\big[\,\log\left(\nt-ir\right)\big]+{\cal O}(1)~,
\end{equation}
where we used Eq.~(\ref{F234}). It is clear that the logarithmic terms
cancel and as a result $\la\phi^2(x)\ra^{ren}_{AF}$ is bounded at
infinity. This is expected, because the bulk is flat. Intuitively,
in four dimensional dS space, the perturbations grow
because when the modes are stretched to a super-horizon scale,
they freeze out. Since modes of ever smaller scales continuously
being stretched, they pile up at a constant rate \cite{vilenkin}.
Since this effect is due to the local curvature of the spacetime,
it should not happen in a flat bulk.

Accordingly, on the brane we recover the same behavior as in de
Sitter space. Indeed, restricting (\ref{g0inf}) on the brane, we
obtain $G_{LC}^{(1)} \sim  \chi$.  We have mentioned before that
$D_{simple}(x,x)$ is UV divergent on the brane. Since this happens
because point splitting regularization used here does not operate
on the brane, this object needs UV regularization and
renormalization. This can be done in a variety of ways, {\em e.g.}
with dimensional regularization
\footnote{
It is
illustrative to consider the massless conformal case. The Green's function
(\ref{mu=0}) for $x$ and $x'$ on the brane is simply
$D^{(ren)}_{\mmu=0}(x,x')\sim 1/|x-x'|^{n/2}$. If $n$ is negative,
this is clearly zero in the coincidence limit, and hence the continuation
to $n=3$ is zero as well. In the non-conformal case, we can use
the integral form (\ref{integral}) to compute
$\phisq\delta(r-r_0)$ in arbitrary dimension. Upon continuation to
$n=3$, one obtains a pole $\sim1/(n-3)$ and a nonzero finite part.
The pole can be absorbed in the brane tension and
$\la\phi^2(r_0)\ra^{ren}$ is a finite constant. },
introducing a finite brane
thickness, smearing the field etc. The point is that the
renormalized value must be a constant simply because
$D_{simple}(x,x)$ is dS invariant and is a function of $x$ only.
Thus, Eqs.~(\ref{gsimpleinf}) and (\ref{g0inf}) imply that on
the brane, $\la\phi^2\ra$ grows in time, as in dS space.

\subsection{Massless minimal coupling}
\label{sub:mmc}
Now, we shall see essentially the same features arise for a
massless minimally coupled scalar in the GK vacuum, in which case
everything can be obtained in a dS invariant way \cite{gaki}.
Because of the shift symmetry $\phi\to\phi+$const. , $\phisq$ is
not an observable in this case. Still, it is possible to define a
`shift-invariant' notion for the field perturbations. Following
\cite{gaki}, one introduces the correlator
\begin{equation}\label{phixy}
{\cal G}(x,y)\equiv\big\langle\left[\phi(x)
-\phi(y)\right]^2\big\rangle_{GK},
\end{equation}
which can be thought of as the combination
$\big[G^{(1)}(x,x)+G^{(1)}(y,y)-2G^{(1)}(x,y)\big]/2$.
Since the first
two terms are UV divergent, we shall rather consider
\begin{equation}\label{phixyren}
{\cal G}^{ren}(x,y)=
{1\over2}\left[G^{ren(1)}(x,x)+G^{ren(1)}(y,y)-2G^{ren(1)}(x,y)\right].
\end{equation}
The main point is that all
the terms of the form $f(x)+f(y)$ in $G^{ren (1)}$ cancel out in
the combination ${\cal G}^{ren}(x,y)$. All the terms in Eqs.~(\ref{g0})
and (\ref{gaf}) that do not vanish in the limit $\alpha\to0$ are
of this form because $\uu^\bs$ is constant in the massless
minimally coupled case. Thus, this correlator is well defined for
the GK vacuum \cite{gaki} and we can readily write
\begin{equation}\label{regsimple}
{\cal G}^{ren}(x,y)={1\over2}\left[
D^{(1)}_{simple}(x,x)+D^{(1)}_{simple}(y,y)-2D^{(1)}_{simple}(x,y)\right]~,
\end{equation}
with $D_{simple}(x,y)$ given by (\ref{gsimple}).\footnote{Note
that (\ref{regsimple}) is completely regular on the light cone.}
Thus, the behavior of the perturbations in the GK vacuum for $x$
and $y$ distant is thus described by the asymptotic behavior of
$D_{simple}(x,x)$ and $D_{simple}(x,y)$. The former is summarized
in Eq.~(\ref{gsimpleinf}).
As mentioned before, for finite $x$, $D_{simple}(x,x)$ is regular
everywhere except on the brane (in which case $y_I=y$).

\begin{figure}[tb]
$$
\begin{array}{cccc}
  \includegraphics[width=4.2cm]{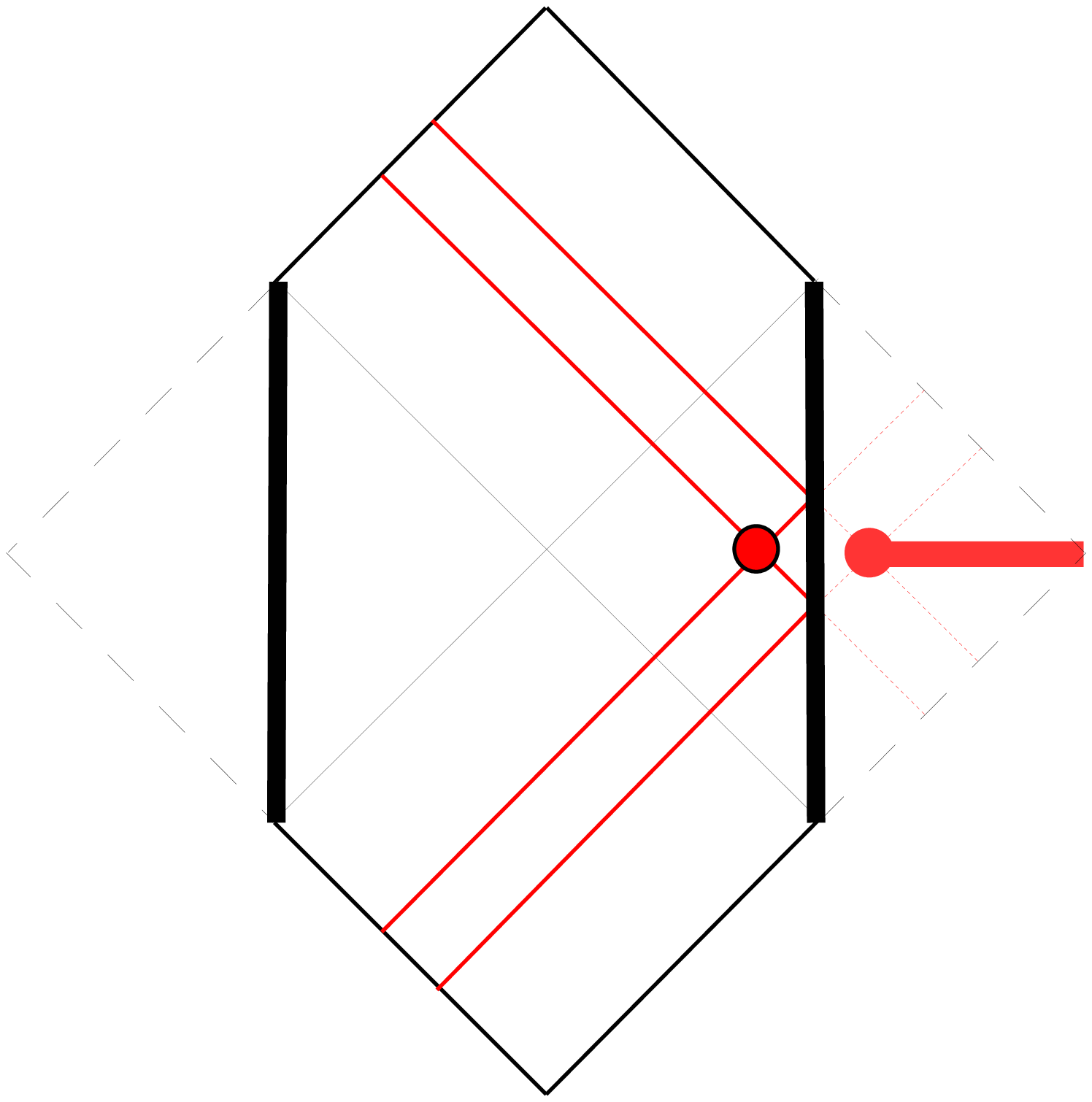}
  &\includegraphics[width=4.2cm]{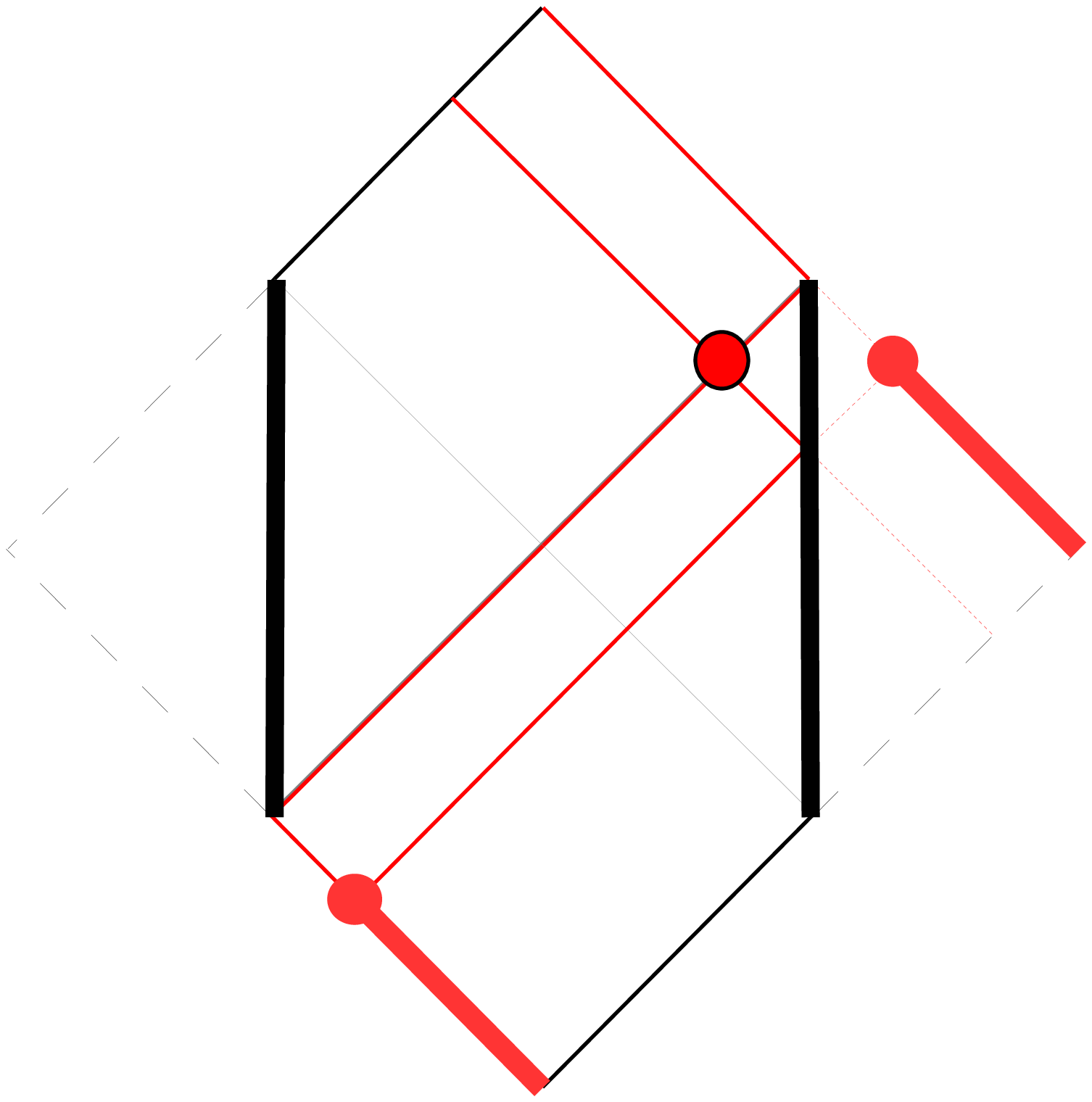}
  &\includegraphics[width=4.2cm]{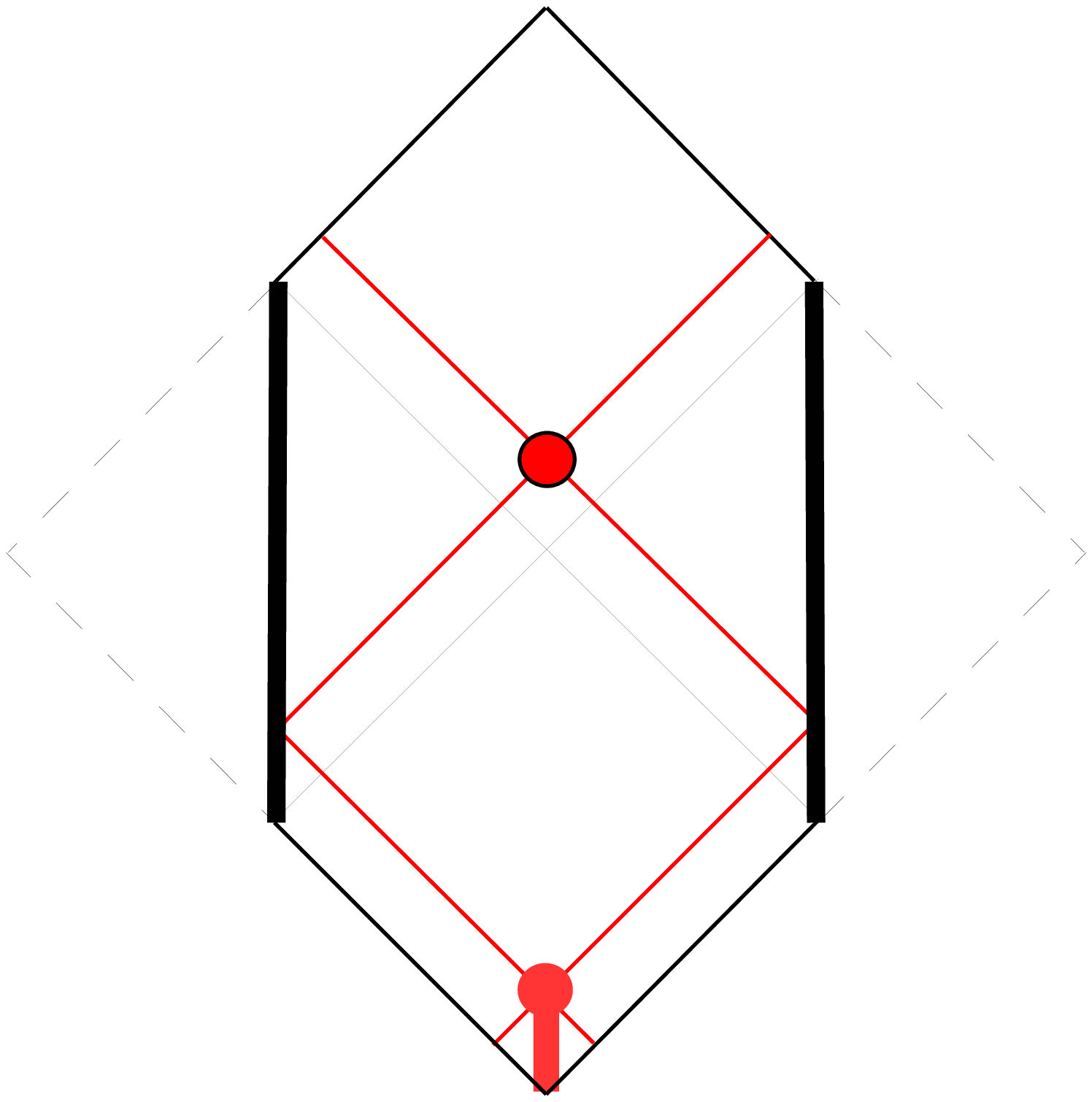}
  &\includegraphics[width=4.2cm]{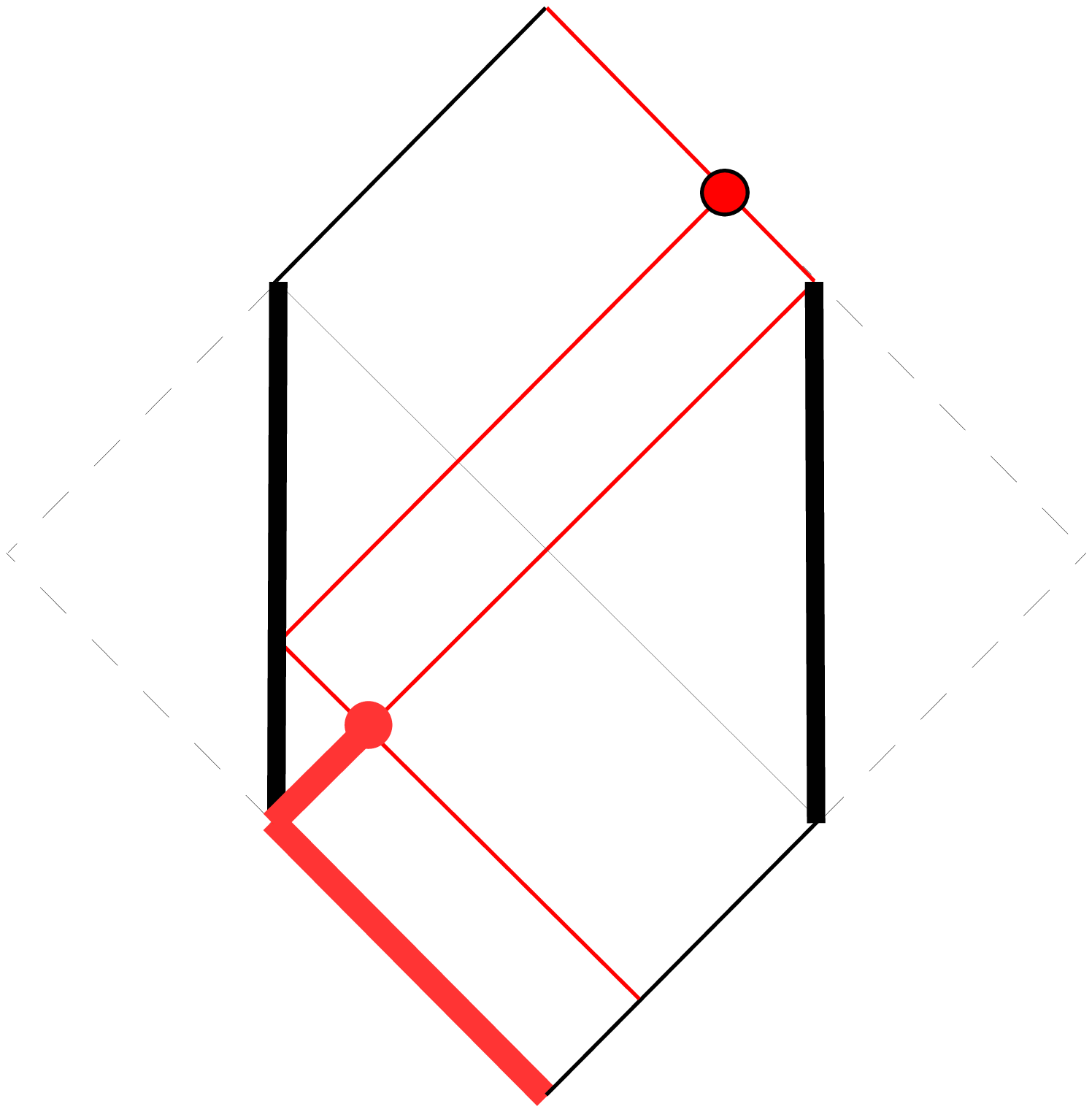}\nonumber\\[-1cm]
  {\rm (a)}& {\rm (b)}& {\rm (c)}& {\rm (d)}
\end{array}
$$
  \caption{ The singularities in $D_{simple}(x,y)$ are located on the light
  cone emanating from $y_I$, the image point of $y$. Also when $x$
  is on the light wedge emanating from `image string' stretching
  from $y_I$ to infinity. Here, we plot the image point
  (smaller filled circle) with the image string attached to it. In
  (a), $y$ is spacelike, in (b) null, in (c) timelike and
  in (d) at null infinity.
  }\label{rays}
\end{figure}

\begin{figure}[htb]
%\begin{eqnarray}
  \includegraphics[width=5cm]{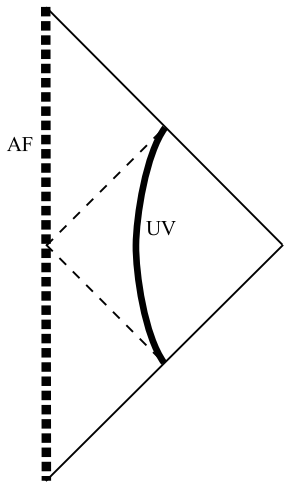}
%&\qquad\qquad
%  \includegraphics[width=5cm]{confDiagram5.eps}
%\nonumber
%\end{eqnarray}
  \caption{  Divergences in $\la\phi^2\ra^{ren}=G(x,x)^{ren}$ and $\tmn^{ren}$ when the bound
state is massless. For the AF vacuum, both the Green's
function and the stress tensor diverge at $\rrho=0$, indicated
with a thick dashed line. In addition, if $\xi\neq0$ the stress
tensor diverges on the light cone like $1/U$ or logarithmically in
four and six dimensions respectively (thin dashed line). Besides,
we also have the ultraviolet `Casimir' divergence on the brane,
represented by the plain thick line. In the Garriga Kirsten vacuum
(for the massless minimally coupled scalar),  $\tmn^{ren}$
presents only this Casimir-type divergence on the brane.}
\label{diagram}
\end{figure}

Before describing the asymptotic form of $D_{simple}(x,y)$,
we need to discuss the singularities that it contains.
The combination that we called $\Delta$ in Eq.~(\ref{gsimple}) is
$$
\Delta(x,y)=|x-y_I|^2/|y_I|^2~,
$$
where $|x|^2=\eta_{\mu\nu}x^\mu x^\nu$ and $y_I^\mu=(r_0^2/|y|^2)
y^\mu$ is the image of $x$ (see comments around
Eq.~(\ref{integral}) and Fig.~\ref{diagram}).
Note that
when $y$ is in one of the Milne regions, then $y_I$ is in the
other one. Rather than an `image charge', it represents the point
where the light cone focuses, (see Fig.~\ref{rays} (b)).
For $x\neq y$ and both finite, $D^{(1)}_{simple}(x,y)$ has
singularities in two types of situations. One is when $x$ is on
the light cone of $y_I$  (then, $\Delta=0$). The other case arise when
the argument of the logarithmic term in Eq.~(\ref{gsimple}) vanishes.
It is convenient to rewrite this term as
\begin{equation}\label{arglog}
-{3\over 8\pi^2 r_0^3}\;
\log\Big|{\left(|x-y_I|+|y_I|\right)^2-|x|^2\over 2|y_I|^2}\Big|
~.
\end{equation}
%Aside from when $x$ coincides or sits on the light cone of $y_I$,
The argument vanishes when they are aligned with respect to the
origin \emph{and} $|x|^2>|y_I|^2$. This condition defines an
`image string' stretching from $y_I$ to infinity.
Hence, this singularity occurs only when two points are in
different Milne regions. This is
consistent with the interpretation of Eq.~(\ref{integral}), that
the Green's function can be constructed with a mirror image and a
linear charge distribution over such a string, see
Fig.~\ref{rays}.\footnote{In four dimensions \cite{xavi}, the
logarithmic term is $\log \big||x-y_I|^2/|y_I|^2\big|$. Thus, this
string singularity does not appear, despite the same
interpretation of Eq.~(\ref{integral}) holds. %In fact, one can show
%that these properties are generic for even and odd dimensions
%respectively.
}
%When $x$ and $y$ are in Rindler region, and
%$x$ is on the `light wedge' of $y_I$ (with cylindric
%constant time $T$ slices), the argument of logarithm also vanishes.
%Here the `light wedge' means the boundary of the causal future or past of the
%image string with $|x|^2>|y_I|^2$.
These two situations correspond to the coincidence of
$x$ with the image of $y$ or its `light cone'.
Therefore all of them are of UV
nature.\footnote{As an aside, let us comment on the singularities
present in the bound state contribution. Specifically, in the term
$\displaystyle
\partial_p \left[(p-p_d) G_p^{D-1}\right]$ present in
Eq.~(\ref{gdisc}). We remind that this term is canceled by the KK
contribution (\ref{gcont}). Hence, it plays no role in the Green's
function. From \cite{gaki,af} we know that for $x$ and $y$
timelike related and far apart, then $\zeta$ is large and pure
imaginary, and it behaves like
$$
%G_{NZM}
\partial_p \left[(p-p_d) G_p^{D-1}\right]\sim \log\big|1-\cos\zeta\big|
= \log\Big|{(|x|-|x_I|)^2-|x-x_I|^2\over2|x||x_I|} \Big|.
$$
This presents divergences on the light cone and whenever $x$ and
$y_I$ (or equivalently, $y$) are aligned. This condition reduces
to the coincidence limit on the brane, but in the bulk one does
not expect singularities to appear at all these points. The above
mentioned cancelation guarantees that these unphysical
singularities are not present in $G^{ren}(x,y)$ once we include
the KK contribution.}

The asymptotic form of $D_{simple}(x,y)$ for distant points
follows from Eq.~(\ref{arglog}) because then, the inverse powers of
$\Delta$ in (\ref{gsimple}) are finite. Taking $y$ fixed and
$x\to\infty$ (with $x$ not on the light cone from $y_I$ nor on the image
string nor on its light wedge), one obtains
\begin{equation}\label{simplexy}
D^{(1)}_{simple}(x,y)\simeq -{3\over 8\pi^2 r_0^3}\;
\log\big|x\big| ~.
\end{equation}
However, since $D_{simple}(x,x)$ grows twice as fast (see
Eq.~(\ref{gsimpleinf})), the combination (\ref{phixyren})
%$$
%{\cal G}^{ren}(x,y)\simeq {1\over2}
%\big[D_{simple}(x,x)-2D_{simple}(x,y)\big]
%$$
is bounded in the bulk.

We can consider as well both $x$ and $y$ approaching null infinity
in the bulk. In this case, the image $y_I$ approaches the light
cone, as illustrated in Fig.~\ref{rays} (d). Then the logarithm in
Eq.~(\ref{arglog}) is $\sim\log\big(|x-y_I|/|y_I|^2\big) \simeq
\log\big(|x|\,|y|\big)$, hence the combination ${\cal G}^{ren}$ is
bounded again.

On the brane, the situation is very different. As mentioned
before, $D_{simple}(x,x)$ is divergent but when properly
renormalized, it is simply a constant because of dS symmetry.
Then, ${\cal G}^{ren}(x,y)$ behaves like $D_{simple}(x,y)$ and
from Eq.~(\ref{arglog}) for large separation and $|x|=|y|=r_0$, one
obtains
$$
{\cal G}^{ren}(x,y)\simeq {3\over 4\pi^2 r_0^3}\;
\log\big|x-y\big| ~.
$$
Since $|x-y|^2=2r_0^2(1-\cos\zeta)$, this corresponds to the
linear growth with the invariant dS time interval
$\zeta(x,y)$ between $x$ and $y$ \cite{gaki}.

%The absence of growth  of perturbations in the bulk can be
%understood as follows. In \cite{gaki}, ${\cal G}$ was used to%
%describe the field perturbations seen by two freely falling
%observers initially nearby. One takes the points $x$ and $y$ to be
%on the world lines of such observers. In dS space, the observers
%separate from each other exponentially fast. After some time,
%their Hubble volumes stop overlapping each other. Before that
%time, the random walks of the field at the two points are
%correlated. But after that they are not, and the field starts
%fluctuating. Here, the same happens on the brane. But in the bulk
%the picture is very different. Two observers moving along parallel
%geodesics close to each other will remain close always, and the
%field correlation will hardly change.

%The behaviour of ${\cal G}(x,y)$ %for distant points
%is somewhat different in the AF vacuum. Due to the crossed terms
%in (\ref{gaf}), in this case we should add to (\ref{regsimple})
%the piece
%$$
%\alpha^2{\uu^\bs(r_x) \uu^\bs(r_y)\over S_{(n)}}
%Re\left[(F_{AF}(x)-F_{AF}(y))^2\right].
%$$
%This is bounded in the Rindler region but is not in the Milne one.
%In fact, whenever $x$ or $y$ approach $\rrho=0$, generically it
%diverges.

Finally, we shall mention that %even though $\tmn$ is ill defined
the generalization of Eq.~(\ref{phixy}),
\begin{equation*}\label{calgmass}
%{\cal G}^{ren}(x,y)&\equiv&
\Big\langle\left[{\phi(x)\over\uu^\bs(x)}
-{\phi(y)\over\uu^\bs(y)}\right]^2\Big\rangle^{ren}%\cr
\end{equation*}
is completely regular in the GK vacuum when $M\neq0$, and its
asymptotic behaviour at infinity parallels that of (\ref{phixy}).
%For $|x|\to\infty$, it behaves like $\sim{(2/
%nS_{(n+1)})}\log{|x-y_I|/ |y_I|}$.

\section{conclusions}
\label{sec:concl}

Our main results can be summarized as follows. In analogy with
what happens in de Sitter (dS) space \cite{af,gaki}, scalar fields
with a massless bound state in the spectrum do not have a well
defined dS invariant vacuum, except for the massless minimally
coupled case. (The case of vanishing bulk mass with non-vanishing
curvature coupling has a little subtlety, though.)
The Green's function and the v.e.v. of the energy
momentum tensor diverges everywhere. The simplest alternative from
the analogy to dS case is to take the Allen Follaci vacuum.
However, in this vacuum, divergences in the stress tensor are not
removed completely within the bulk. Figure \ref{diagram}
illustrates the location of the IR singularities in the Green's
function and the stress tensor for the AF vacuum. It remains to
clarify whether it is possible or not to avoid these singularities
by choosing a vacuum for the KK modes other than the
dS invariant vacuum.

When the bound state is very light (but not exactly massless)
because $M$, $\mbr$ and/or $\xi$ are fine tuned according to
Eq.~(\ref{muc}), then the  stress tensor in dS invariant
vacuum takes
the form of (\ref{IRdivtmn}). The stress tensor in this case is
smooth, but it becomes very large. Hence, even when the bound
state mass is not exactly zero, the dS invariant vacuum looks
problematic because of large back reaction. Note that the
situation here is different from the usual dS case in two
respects. In dS space the large v.e.v. in the stress tensor for
the dS invariant vacuum is a constant proportional to the metric.
Hence, it might be absorbed by IR renormalization of the
cosmological constant. In our case, the stress tensor given by
Eq.~(\ref{IRdivtmn}) is a nonlocal expression and cannot be
`renormalized away'. On the other hand, if one does not want to
make any IR renormalization, in the dS case one can take the AF
vacuum, and $\tmn$ stays regular. In the brane world, we do not
know the prescription how to remove this large v.e.v. by changing the
vacuum state. Choosing non dS invariant vacuum will lead to not
only the mentioned divergence
in the bulk, but also a new singularity on the light cone,
when the bound state
mass $m_d$ is not exactly zero. In this case, the radial function
for the bound state behaves like $\propto r^{-m_d^2/n}$ near the
light cone at $r=0$. Hence if you single out the contribution to
the Green's function from the bound state, it is singular and its
derivatives diverges at $r=0$. Hence, as far as we restrict the
change of quantum state to the bound state, we will not be able to
remove the large v.e.v. of stress tensor without spoiling its
regularity.

%Here we would like to mention that the above features are quite
%different from those in the case of ordinary dS space. In dS case,
%the large v.e.v. in $\tmn$ in dS invariant vacuum is a constant
%proportional to the metric. Hence, it might be absorbed by IR
%renormalization of the cosmological constant. Even if we do not
%accept this rather hand-waving IR renormalization, vacuum state
%analogous to the AF vacuum can remove the large v.e.v. without
%spoiling the regularity.

%Note that (\ref{IRdivtmn}) is a nonlocal
%expression (except for $M=0$), so it cannot be renormalized away.
%In the special case $M=0$, the bulk contribution from this mode
%vanishes and the brane term is of the form of a brane tension. In
%the RSII model, the bulk contribution from the bound state
%vanishes when $M^2+\xi {\cal R}=0$
%Even if we have a
A light bound state is compatible with a well behaved and not
large stress tensor only in a situation `close' to the massless
minimally coupled case. More precisely, here we consider the cases
that all of the bulk mass $M$, the brane mass $\mbr$ and the
nonminimal coupling $\xi$ are small (see Eq.~(\ref{BSmassMsmall})).
This corresponds to having a light bound state without accidental
cancellations. Namely,  the squared bound state mass $m_d^2$ is of
order of the largest among $M^2$, $H \mbr$ and $H^2 \xi$, where
$H$ is the Hubble constant on the brane. In this case, the dangerous
terms proportional to $m_d^{-2}$ are always associated with some small
factor $M^2$, $\mbr$ or $\xi$, and therefore none of them becomes
large.

The case with $M^2\approx H\mbr \ll H^2\xi$ has a little subtlety.
In this case, the large v.e.v. in $\tmn$ appears only in
the brane part, and it is constant proportional to the
induce metric. Hence, we might be able to consider a model with
an appropriate IR renormalization. Only in such a modified model,
the stress tensor in the dS invariant vacuum can
escape from appearance of a large v.e.v..

Application to the bulk inflaton type models
\cite{kks,hs} is a part of motivation of the present study.
In these models there must be a light bound state of a
bulk scalar field.
In order to explain the smallness of the bound state mass
it will be natural to assume that it is
due to smallness of all the bulk and brane
parameters without fine tuning.
Therefore we will not have to seriously worry about
the backreaction of the inflaton in the context of bulk inflaton
type models.

We have a few words to add on the massless minimally coupled case.
If we consider this case as a limiting situation
close to the massless minimally
coupled case, the results depend on how we fix the ratio amongst
$M$, $H \mbr$ and $H^2 \xi$, and hence there remains ambiguity.
However, this limiting case has the
shift symmetry $\phi\to\phi+$constant. If this shift symmetry is
one of the symmetries that are to be gauged, there is no ambiguity
because the problematic homogeneous mode does not exist in the
theory from the beginning. In this setup, undifferentiated $\phi$
is not an observable. In fact, $\tmn$ automatically does not
contain undifferentiated $\phi$. Hence, $\tmn$ is unambiguously
defined although the Green's function is not well defined. We can
compute $\tmn$ in this model by applying the idea of the Garriga
Kirsten vacuum (equivalent to a limiting case of AF vacuum), and
we confirmed it dS invariant and regular as is expected.

We have also discussed the form of the field perturbations
$\phisq$ when the bound state is massless. The main point is that
on the brane $\phisq$ grows linearly with dS time $\chi$ while in
the bulk it is bounded, as expected since the bulk is flat. Aside
from this, we derived in closed form the $\tmn$ for a generic
field with zero bulk mass, see Eq.~(\ref{tmassless}).

The same discussion applies in the RSII model with little
modifications, which mainly comes from the fact that
the Ricci tensor term in the bulk stress
tensor $\sim \xi\phisq {\cal R}_{\mu\nu}$ is not zero in the
RSII model. Thus,
the bulk part of $\tmn$ is finite in the limit of massless bound
state only for massless minimal coupling.
%Also, the perturbations
%are expected to grow on the brane and to be bounded in the bulk.

\acknowledgements

We are grateful to Jaume Garriga, Misao Sasaki, Wade Naylor, Alan
Knapman and Luc Blanchet for useful discussions. T.T. acknowledges
support from Monbukagakusho Grant-in-Aid No. 16740141 and Inamori
Foundation, and O.P. from JSPS Fellowship number P03193.

\appendix

\section{Massive Green's function in de Sitter}
\label{sec:GdS}

In this Appendix we obtain the form of the Green's function for a
massive field propagating in $D-1=n+1$ dimensional de Sitter (dS)
space.
The metrics for dS space and its Euclidean version are
\begin{equation}
 dS_{(n+1)}^2
= -d\chi ^2+ \cosh^2\chi  d\Omega_{(\q)}^2 = d\ce ^2+\sin^2\ce
d\Omega_{(\q)}^2,
\end{equation}
where $d\Omega_{(n)}^2$ is the metric of a unit $n$ dimensional
sphere. The Euclidean time is given by $\ce= i\chi + \pi/2$. The
Euclidean version of the Hadamard  function in dS space can
be found from the equation
\begin{equation}
 \left[\partial_{\ce}^2+\q\cot\ce \partial_{\ce} -
     \left(p^2+{\q^2\over 4}\right)\right]G^{(dS)}_p(x,x')
=-\delta^{(\q+1)}(x-x').
\end{equation}
From the symmetry, we can choose $x'$ to be at the pole so that
$G^{(dS)}_p$ depends on $\ce$ only. In terms of  $F=(\sin(\ce
-\pi))^{(\q-1)/2} G^{(dS)}_p(x,x')$, this becomes
\begin{equation}
 \left[(1-w^2)\partial_w^2-2w\partial_w -\left(p^2+{1\over 4}\right)
       -{(\q-1)^2\over 4 (1-w^2)}\right]F=0,
\end{equation}
where $w=\cos(\ce -\pi)$, and this is solved to give
\begin{equation}
 F=N e^{(\q-1)\pi i\over 2}\Gamma\left({\q+1\over 2}\right)
      P^{-{\q-1\over 2}}_{ip-{1\over 2}}
     (\cos(\ce -\pi)).
\end{equation}
Hence, the Green's function will be given by the replacement of
$\ce$ by the proper distance between the points $x$ and $x'$ in
$dS$ space, which we call $\zeta(x,x')$. $G^{(dS)}_p$ is
guaranteed to be regular at $\zeta \to \pi$. To see the behaviour
in the $\zeta \to 0$ limit, the alternative expression
\begin{eqnarray}
\label{gds}
 G^{(dS)(1)}_p&=&{\tilde N\over (1-\cos\zeta)^{\q-1\over 2}}
  \Biggl\{F\left(-ip+{1\over 2},ip+{1\over 2},{-\q+3\over 2};
     {1-\cos\zeta\over 2}\right)\cr &&+
     {\Gamma\left({\q\over 2}-ip\right)\Gamma\left({\q\over 2}+ip\right)
          \Gamma\left(-{\q-1\over 2}\right)
     \over \Gamma\left({1\over 2}-ip\right)\Gamma\left({1\over 2}+ip\right)
          \Gamma\left({\q-1\over 2}\right)
      }\left({1-\cos\zeta\over 2}\right)^{\q-1\over 2}
      F\left(-ip+{\q\over 2},ip+{\q\over 2},{\q+1\over 2};
     {1-\cos\zeta\over 2}\right)
   \Biggr\}
\end{eqnarray}
is relevant, and here
\begin{equation}
  \tilde N=
      {\Gamma\left({\q+1\over 2}\right)\Gamma\left({\q-1\over 2}\right)
      \over \Gamma\left({\q\over 2}-ip\right)\Gamma\left({\q\over
      2}+ip\right)}N~.
\end{equation}
In the coincidence limit, the Green's function must behave like
$G_{(n+1)}^{(1)}\approx 1/(n-1) S_{(\q)}\zeta^{\q-1}$, where
$S_{(\q)}$ is the area of an $\q$-dimensional unit sphere. The
limiting behaviour is controlled by the first term. Hence we have
$\tilde N=2^{-(\q-1)/2}/(n-1)S_{(\q)}$.

\section{Massless Green's function in dS}
\label{sec:GdSmassless}

Here, we compute the Green's function for the massless scalar in
dS. In the massless limit ($p\to in/2$), the de Sitter (dS) invariant
Green's
function diverges because of the contribution from the $\ell=0$
mode, ${\cal Y}_{p00}$. The idea is to construct a modified
Green's function by substituting this mode by another one that is
finite in the limit $p\to in/2$. In other words,
$$
G_{(m=0)}^{(dS)(+)}=G_{(\ell>0)}^{(+)} + {\widetilde {\cal
Y}}_{}^{}{\widetilde {\cal Y}_{}}'^{*},
$$
where ${\widetilde {\cal Y}}$ corresponds to the $\ell=0$ mode and
$$
G_{(\ell>0)}^{(+)}=\sum_{\ell>0} {\cal Y}_{p\ell m}(\chi) {\cal
    Y}^*_{p\ell m}(\chi')~,
$$
where ${\cal Y}_{p\ell m}$ are the positive frequency
dS invariant vacuum modes. The latter can be obtained as follows. Since
$G^{(dS)}_p$ diverges because of the $\ell=0$ mode,  the
divergent term in the Laurent expansions of both $G^{(dS)}_p$ and
${\cal Y}_{p00}{\cal Y}_{p00}'^*$ coincide. We show below that
this series contains a simple pole in $p-in/2$. Hence, we can
write
\begin{equation}\label{l>0}
G_{(\ell>0)}^{(+)}=\lim_{p\to{in\over2}} \left[ G^{(dS)(+)}_p -
{\cal Y}_{p00}^{~} {\cal Y}_{p00}^{*}\right] =\partial_{p}\left[
\left(p-i{n\over2}\right) G^{(dS)(+)}_p \right]_{p=i{n\over2}} -
\partial_{p}\left[\left(p-i{n\over2}\right) {\cal Y}_{p00}^{~}
{\cal Y}_{p00}'^{*} \right]_{p=i{n\over2}} ~.
\end{equation}
The explicit expression for the first term in the r.h.s. is
unnecessary because it is canceled by the KK contribution
(\ref{gcont}). The dS invariant vacuum mode
for $\ell=0$ with $-ip<n/2$ is
\begin{equation}\label{y0}
{\cal Y}_{p00}=C_p \; { 2^{\q-1\over2}\Gamma\left({\q+1\over
2}\right)\over \left(\cosh\chi\right)^{\q-1\over2}}
        \;P^{-{\q-1\over 2}}_{-ip-{1\over 2}}(i\sinh \chi)~,
\end{equation}
and the normalization constant has been chosen so  that
$\lim_{p\to in/2}\, {\cal Y}_{p00}/C_p=1$. The Klein Gordon
normalization requires
\begin{equation}\label{cp}
|C_p|^2= {\Gamma\left(ip +{\q\over 2}\right)
  \Gamma\left(-ip +{\q\over 2}\right)
  \over 2^\q \Gamma\left({\q+1\over 2}\right)^2 S_{(n)}}~.
\end{equation}
Expanding around $p=in/2$ we find
\begin{equation}\label{cpsmall}
|C_p|^2={1\over i n S_{(n+1)} } {1\over p-in/2} + {\cal
O}\left[(p-in/2)^0\right]~,
\end{equation}
where $S_{(n+1)}$ is the area of an $n+1$ dimensional sphere of
unit radius. The behaviour of ${\cal Y}_{p00}$ near $p=in/2$, is
most easily found using the equation that it solves,
\begin{equation}
 \left[\partial_\chi^2+\q\tanh \chi\partial_\chi +
     \left(p^2+{\q^2\over 4}\right)\right] {\cal Y}_{p00} =0~.
\end{equation}
Bearing in mind that the positive frequency function for the
dS invariant vacuum is determined by the regularity when the function
is continued to the Euclidean region on the side that contains
$\chi=-\pi i/2$, one finds
\begin{equation}
\label{yp00} {\cal
Y}_{p00}(\chi)=C_p\;\left[1-\left(p^2+{\q^2\over 4}\right)
   \int^\chi d\chi_{{}_1}{1\over \cosh^\q \chi_{{}_1}}
         \int_{-\pi i/2}^{\chi_{{}_1}} d\chi_{{}_2} \cosh^\q
         \chi_{{}_2}
    + {\cal O}\left( \left(p-i n/ 2\right)^2\right) \,\right]~.
\end{equation}
From this result, the last term in Eq.~(\ref{l>0}) can be readily
evaluated, and Eq.~(\ref{g0}) follows.

\section{Divergence in the Green's function and the $\ell=0$ mode}
\label{sec:cancelation}

Here, we show that the IR divergence in the Green's function
(\ref{polesmass}) or (\ref{polesmassless}) is due to the
homogeneous $\ell=0$ mode of the bound state. Using
Eqs.~(\ref{wavefunct}), (\ref{radialnorm}), (\ref{cpsmall}) and
(\ref{yp00}), the contribution from the $\ell=0$ mode of the bound
state for $p_d$ close to $in/2$ is found to be
\begin{eqnarray}
\label{zerozero}
 G_{\ell=0,p_d}^{(1)}&=& \uu^\bs(r)\uu^\bs(r')\;
    {\cal Y}_{p_d00}(\chi) {\cal Y}^*_{p_d00}(\chi') +{\rm c.c.}
\simeq 2\uu^\bs(r)\uu^\bs(r')\;|C_{p_d}|^2  \left[ 1+ {\cal
O}\left( p_d-in/2\right) \right]    \cr
    &\simeq&
    {1\over  S_{(n+1)} } \;{1\over p_d-in/2}\;
    {  I_{n/2}(Mr) I_{n/2}(Mr')\, /\,(r r')^{\q/2}
        \over  I_{n/2}(Mr_0) \left.\partial_p
          \left(\mmu_c I_{-ip}(Mr_0) -M r_0I'_{-ip}(Mr_0)
                \right)\right\vert_{p=in/2}} +\dots
%    &=&
%    %{T_{p_d}(\chi)T^*_{p_d}(\chi')+T_{p_d}(\chi')T^*_{p_d}(\chi)
%{\Gamma\left(ip_d+{\q\over 2}\right)
%               \Gamma\left(-ip_d+{\q\over 2}\right)
%\over 2^{\q} \Gamma\left({\q+1\over 2}\right)^2 S_{(\q)}}
%    { 2p_d (r r')^{-\q/2} I_{-ip_d}(Mr) I_{-ip_d}(Mr')
%        \over  I_{-ip_d}(Mr_0) \left.\partial_p
%          \left(\mmu I_{-ip}(Mr_0) -M r_0I'_{-ip}(Mr_0)
%                \right)\right\vert_{p=p_d}} +\dots
\end{eqnarray}
where the dots denote higher order in $p_d-in/2$. In the same
limit, $\mmu$ is close to $\mmu_c$. The $j=k=0$ term in
(\ref{polesmass}) is
\begin{eqnarray}
\label{cancelation}
  G_{j=0,k=0}^{(ren)(1)}
    &=&  -{ \,n \Gamma\left(\q\right)
         \over 2^{\q}\Gamma\left({\q+1\over 2}\right)^2  S_{(\q)}}
          \;{\mmu K_{\q/2}(Mr_0) -Mr_0 K'_{\q/2}(Mr_0)
           \over \mmu I_{\q/2}(Mr_0) -Mr_0 I'_{\q/2}(Mr_0)}
    \;{I_{\q/2}(Mr) I_{\q/2}(Mr')\over (rr')^{\q/2}}\cr
    &=&  -{ 1 \over  S_{(\q+1)}}
          \;{\mmu_c K_{\q/2}(Mr_0) -Mr_0 K'_{\q/2}(Mr_0) + \left(\mmu-\mmu_c\right) K_{\q/2}(Mr_0)
           \over  I_{\q/2}(Mr_0)\left(\mmu-\mmu_c\right) }
    \;{I_{\q/2}(Mr) I_{\q/2}(Mr')\over (rr')^{\q/2}}\cr
    &\simeq& - { 1 \over  S_{(\q+1)}}
          \;{ I_{\q/2}(Mr) I_{\q/2}(Mr')/ (rr')^{\q/2}
           \over
    I^2_{n/2}(Mr_0) \left(\mmu -\mmu_c \right)} + {\cal~O}\left[(\mmu-\mmu_c)^0\right]
\end{eqnarray}
where in the second line we used Eq.~(\ref{muc}) and the Wronskian
relation $K_{n/2}^{}(z) I_{n/2}'(z)-K_{n/2}'(z)
I_{n/2}^{}(z)=1/z$.
%where we used the relation
%\begin{eqnarray}
% \lim_{\mmu\to\mmu_c} &&\mmu K_{-\q/2}(Mr_0) -Mr_0
% K'_{-\q/2}(Mr_0)\cr
%  &&=  {Mr_0\over I_{\q/2}(Mr_0)}
%   \left[
%    I'_{\q/2}(Mr_0)K_{-\q/2}(Mr_0) -I_{\q/2}(Mr_0) K'_{-\q/2}(Mr_0)\right]
%  =   {1\over I_{\q/2}(Mr_0)}.
%\end{eqnarray}
Using Eq.~(\ref{BSmassM}), it is easy to see that
\begin{eqnarray}\label{taylor}
%    \mmu I_{\q/2}(Mr_0) -Mr_0 I'_{\q/2}(Mr_0)&=&
    \mmu-\mmu_c
=%        &=&
        \partial_{p_d} \mmu
        \big|_{p_d=in/2}\,
        \left(p_d-i{n\over2}\right) + \dots %\cr
=%        &=&
        -{\partial_{p}\left[ \mmu_c I_{-ip} -Mr_0 I_{-ip}'(Mr_0)
        \right]\big|_{p=in/2}\over I_{\q/2}(Mr_0) }\; \left(p_d-i{n\over2}\right) +
        \dots~,
\end{eqnarray}
so (\ref{zerozero}) and (\ref{cancelation}) agree in the limit
$p_d\to in/2$. Note that since in this limit this is the dominant
contribution and $m_d\simeq in(p_d-in/2)$, the total Green's
function (\ref{polesmass}) can be rewritten in the simple form
\begin{equation}\label{glimit}
G^{(1)}_{(ren)}(x,x')\simeq{2\over~S_{(n+1)}}\;{\uu_0^\bs(r)\uu_0^\bs(r')\over
r_0^2\, m_d^2}+{\cal O}\left(m_d^0\right)~,
\end{equation}
where $\uu_0^\bs(r)=N_0 I_{n/2}(r)/r^{n/2}$ is the wave function
of the bound state for the exactly massless case.

\end{document}